\documentclass[11pt,a4paper]{article}
\pdfoutput=1

\usepackage{jcappub}
\usepackage{mathrsfs}
\usepackage{bbm}
\usepackage{bm}
\usepackage{dsfont}
\usepackage{tensor}
\usepackage{xspace}
\usepackage{aeguill}
\usepackage{wasysym}
\usepackage{microtype}
\usepackage{empheq}
\usepackage{ifthen}
\usepackage{amsmath}
\usepackage{subcaption}
\usepackage{wrapfig}

\newcommand\ph{\ensuremath{\varphi}}
\newcommand\eps{\ensuremath{\varepsilon}}

\newcommand{\cst}{\mathrm{cst}}

\newcommand\define{\equiv}

\newcommand\vect[1]{\boldsymbol{#1}}

\newcommand\e[1]{_{\text{#1}}}

\newcommand\U[1]{\:\mathrm{#1}}

\newcommand{\dd}{\mathrm{d}}
\newcommand{\Dd}{\mathrm{D}}

\newcommand{\ddf}[3][]{\frac{\dd^{#1} #2}{\dd {#3}^{#1}}}
\newcommand{\Ddf}[3][]{\frac{\Dd^{#1} #2}{\dd {#3}^{#1}}}

\renewcommand\lim[2]{\underset{ #1 \rightarrow #2 }{ \mathrm{lim} } \,}

\newcommand{\delimiters}[4][]{
\ifthenelse{ \equal{#1}{1} }{  #2 #3 #4  }
					{ \ifthenelse{\equal{#1}{2}}{ \big#2 #3 \big#4 }
						{ \ifthenelse{\equal{#1}{3}}{ \Big#2 #3 \Big#4 }
							{ \ifthenelse{\equal{#1}{4}}{ \bigg#2 #3 \bigg#4 }
								{ \ifthenelse{\equal{#1}{5}}{ \Bigg#2 #3 \Bigg#4 }
									{ \left#2 #3 \right#4 }
								}
							}
						}
					}
													}

\newcommand{\pa}[2][]{\delimiters[#1]{(}{#2}{)}}
\newcommand{\pac}[2][]{\delimiters[#1]{[}{#2}{]}}

\newcommand{\abs}[2][]{\delimiters[#1]{|}{#2}{|}}
\newcommand{\ev}[2][]{\delimiters[#1]{\langle}{#2}{\rangle}}

\newcommand{\tidal}{\mathcal{R}}
\newcommand{\jacobi}{\mathcal{D}}
\newcommand{\wronski}{\mathcal{W}}

\newcommand{\Ricfoc}{\mathscr{R}}
\newcommand{\Weylfoc}{\mathscr{W}}

\newlength{\boxtitlelength}
\newlength{\halfrulelength}
\newcommand{\boxtitle}[1]{\footnotesize\bf{\:#1\:}}

\definecolor{blue4}{RGB}{0,0,143}
\definecolor{red4}{RGB}{143,0,0}
\definecolor{orange}{RGB}{255,128,0}
\definecolor{darkcyan}{RGB}{0,128,128}
\definecolor{olive}{RGB}{0,128,0}
\definecolor{purple}{RGB}{128,0,128}
\definecolor{cyan2}{RGB}{0,255,255}
\definecolor{fushia}{RGB}{255,0,255}
\definecolor{mygray}{gray}{0.5}
\definecolor{lightgray}{gray}{0.85}

\newcommand{\apj}{ApJ}
\newcommand{\apjs}{ApJS}
\newcommand{\apjl}{ApJL}
\newcommand{\aap}{A{\&}A}

\newcommand{\mnras}{MNRAS}

\newcommand{\jcap}{JCAP}

\newcommand{\prd}{Phys. Rev. D}

\newcommand{\nat}{Nature}

\newcommand{\varDA}{\sigma_{D\e{A}}^2}

\makeatletter
\def\@fpheader{\relax}
\makeatother

\title{Ray tracing and Hubble diagrams in post-Newtonian cosmology}

\author[a]{Viraj A A Sanghai,}
\author[b]{Pierre Fleury,}
\author[a]{Timothy Clifton}

\affiliation[a]{School of Physics and Astronomy, Queen Mary University of London,\\
327 Mile End Road, London E1 4NS, United Kingdom}
\affiliation[b]{D\'{e}partment de Physique Th\'{e}orique, Universit\'{e} de Gen\`{e}ve,\\
24 quai Ernest-Ansermet, 1211 Gen\`{e}ve 4, Switzerland}

\emailAdd{v.a.a.sanghai@qmul.ac.uk}
\emailAdd{pierre.fleury@unige.ch}
\emailAdd{t.clifton@qmul.ac.uk}

\abstract{On small scales the observable Universe is highly inhomogeneous, with galaxies and clusters forming a complex web of voids and filaments. The optical properties of such configurations can be quite different from the perfectly smooth Friedmann-Lema\^{i}tre-Robertson-Walker (FLRW) solutions that are frequently used in cosmology, and must be well understood if we are to make precise inferences about fundamental physics from cosmological observations. We investigate this problem by calculating redshifts and luminosity distances within a class of cosmological models that are constructed explicitly in order to allow for large density contrasts on small scales. Our study of optics is then achieved by propagating one hundred thousand null geodesics through such space-times, with matter arranged in either compact opaque objects or diffuse transparent haloes. We find that in the absence of opaque objects, the mean of our ray tracing results faithfully reproduces the expectations from FLRW cosmology. When opaque objects with sizes similar to those of galactic bulges are introduced, however, we find that the mean of distance measures can be shifted up from FLRW predictions by as much as $10\%$. This bias is due to the viable photon trajectories being restricted by the presence of the opaque objects, which means that they cannot probe the regions of space-time with the highest curvature. It corresponds to a positive bias of order $10\%$ in the estimation of $\Omega_{\Lambda}$ and highlights the important consequences that astronomical selection effects can have on cosmological observables.}


\date{\today}

\begin{document}

\maketitle
\flushbottom

\section{Introduction}
\label{sec:introduction}

The Universe we see around us appears to be close to homogeneous and isotropic on large scales, but is very inhomogeneous and anisotropic on small scales. The standard approach to modelling this situation is to assume the existence of a Friedmann-Lema\^{i}tre-Robertson-Walker (FLRW) geometry that obeys Einstein's equations for some averaged matter density. Using this solution as a background, one can then perform a perturbative expansion of both the metric and the energy-momentum tensor, in order to incorporate inhomogeneous astrophysical structures and their gravitational fields. A possible drawback of using this approach is that it assumes that the evolution and averaging operations commute. This is not true in Einstein's equations, where we have terms such as
\begin{equation}
\vect{R} [\bar{\vect{g}}] \neq \bar{\vect{R}}[\vect{g}] \, ,
\end{equation}
due to the non-linearity of the Ricci tensor. This non-commutativity means that the original solution, used to model the background, may not stay a faithful description of the Universe on large scales, even if it started out as such at early times. In other words, there is the possibility in Einstein's theory that small-scale structures could influence the large-scale cosmic expansion; a possibility referred to as ``back-reaction'' in the cosmology literature~\cite{Buchert:1999er,Buchert:2011sx,Clarkson:2011zq,Clifton:2013vxa}.

One way to side-step the difficult mathematical problems associated with back-reaction is to construct bottom-up cosmological models in which the large-scale properties of the Universe emerge, rather than being specified from the beginning. Such models do not require averaging, in the way that is necessary in the standard top-down approach, and instead allow the expansion of the Universe to emerge naturally from the boundary conditions between neighbouring astrophysical bodies. A class of such models that has recently been developed by two of us has been dubbed ``post-Newtonian cosmology'', due to the use of the well-known post-Newtonian expansions in its construction~\cite{Clifton:2010fr,2015PhRvD..91j3532S, 2016PhRvD..94b3505S}. These models are described in more detail in section~\ref{subsec:PN_cosmo} below, and will form the basis of the study of cosmological redshifts and distance measurements that we perform in this paper. They are of particular interest for the study of ray optics that we perform here as they naturally allow for the presence on non-linear structures in a fully self-consistent and unambiguous way.

While the kinematical back-reaction effects in these models have been precisely quantified in refs.~\cite{2015PhRvD..91j3532S,2016PhRvD..94b3505S}, the optical properties have not until now been explicitly calculated. These latter properties are of great practical importance, as they are the direct observables upon which almost all astronomical probes are based. Above and beyond any questions involving the large-scale expansion, the influence of structure on the optical properties of a space-time are of very significant interest for the determination of cosmological parameters. If non-linear structures have any systematic effect on the propagation of rays of light, then this could potentially have significant influence on any inferences of redshifts and distance measures over cosmological scales, and could consequently bias the estimation of (for example) the amount of dark energy in the Universe. The appropriate formalism for investigating the optical properties of the Universe is geometric optics, which is outlined in section~\ref{subsec:light_propagation_curved_space-time}. We will apply this formalism to our post-Newtonian cosmological models, using direct ray tracing techniques, in order to determine the influence of non-linear structure on observables such as the distance-redshift relation. 

Of course, the optical properties of inhomogeneous cosmological models have been extensively studied in the past, and various frameworks have been developed to try to model the general behaviours that are expected from the effects of a lumpy matter content. Of particular relevance for the present study are the Einstein-Straus Swiss cheese models in which regions of FLRW geometry are excised and replaced by Schwarzschild~\cite{1945RvMP...17..120E,1946RvMP...18..148E,1969ApJ...155...89K}, the Lindquist-Wheeler models that construct an approximate space-time out of Schwarzschild directly~\cite{1957RvMP...29..432L}, and the Bruneton-Larena model that creates an approximate model that is valid for a short period of cosmic time~\cite{2012CQGra..29o5001B}. The optical properties of these models have all been studied in the past, and we will refer to them further below. In particular, the post-Newtonian cosmological models can be considered an improved version of the Lindquist-Wheeler model as the approximation scheme is under much better control. It could also be considered an improvement on the Swiss cheese and Bruneton-Larena models, as it removes the need for a background FLRW geometry and is valid for much longer periods of cosmic time.

In section~\ref{sec:fundamentals} we discuss the physics of our cosmological model, and the optical equations necessary to calculate redshifts and distance measures. In section~\ref{sec:method} we then outline the method we use to calculate the optics along very many lines of sight. Section~\ref{sec:results} then gives a detailed account of the results of our numerical integrations. This is followed in section~\ref{sec:discussion} by a discussion of these results in the context of the theorems of Weinberg, Kibble \& Lieu, the stochastic approach to lensing, and the recent results that have been obtained using methods from numerical relativity. Finally, in section~\ref{sec:conclusion} we conclude. We work with geometrised units throughout, where~$G=c=1$. Bold symbols can refer to four-vectors, spatial vectors, or matrices. Latin indices~$a,b,c,\ldots$ are space-time indices and run from 0 to 3. Greek indices~$\mu,\nu,\rho,\ldots$ are spatial indices and run from 1 to 3. We use the signature $(-+++)$ for the metric, and the gravitational potentials~$\Phi,\Psi$ are taken to be positive. Appendices~\ref{app:geometry} and~\ref{app:tests} give details of the components of geometric quantities within our perturbative expansion, and the numerical tests we performed on our code in de Sitter space-time.

\section{Fundamentals}
\label{sec:fundamentals}

This section introduces the key ingredients for the present work; namely the post-Newtonian cosmological models with non-linear matter fields developed in refs.~\cite{2015PhRvD..91j3532S,2016PhRvD..94b3505S}, and the laws of geometric optics in curved space-times as discussed in refs.~\cite{1965gere.book..249P,1992grle.book.....S,2004LRR.....7....9P,2015arXiv151103702F}.

\subsection{Post-Newtonian cosmology}
\label{subsec:PN_cosmo}

The post-Newtonian cosmological models are built block-by-block, by patching together regions of post-Newtonian space-time \cite{Clifton:2010fr}. The large-scale dynamics of the resulting universe then emerges naturally from the junction conditions at the boundaries between these regions. The purpose of such a model is twofold. Firstly, it allows us to build a cosmological solution constructively, from the bottom up. This means that the expansion of such a universe does not need to be assumed from the outset, as is the case in most approaches to cosmological modelling, and that the effects of inhomogeneity on the expansion dynamics can be studied in an unambiguous way, without assuming anything about averaging or coarse-graining. Secondly, it makes explicit the link between the gravitational fields of the objects that exist within the Universe and the expansion of the Universe itself. This allows us, for example, to construct self-consistent parametrized frameworks that can be used to test gravity in both cosmology and the weak-field limit \cite{Sanghai:2016tbi}. It also allows us to build Hubble diagrams from the gravitational fields associated with non-linear astrophysical bodies themselves, rather than in any putative averaged ``background''. It is this last possibility that we study here.

\subsubsection{The post-Newtonian formalism}

The post-Newtonian formalism~\cite{will1993theory} is a perturbative expansion scheme for the equations of general relativity, which is intended to deal with weak gravitational fields generated by slowly moving sources. The expansion itself is performed in terms of the smallness parameter
\begin{equation}
\eps \define \frac{v}{c} \ll 1 \, ,
\end{equation}
where $v$ is the typical velocity of sources of the gravitational field and $c$ is the speed of light. This implies that, for consistency, the typical time derivative of any field~$\ph$ must be much smaller than its typical spatial derivative
\begin{equation}
\partial_t \ph \sim \eps \partial_\mu \ph \ll \partial_\mu \ph \, .
\end{equation}
The space-time metric and the matter fields are then expanded using this rule, and the field equations are constructed and solved, order-by-order in $\eps$. For the metric this means writing~$g_{ab}=\eta_{ab}+h_{ab}$, where $\vect{\eta}$ denotes the Minkowski metric and $\vect{h} \ll \vect{\eta}$ is a perturbation, so that the line-element becomes
\begin{equation}
\dd s^2 = \pac{-1 + h_{tt}^{(2)}+ h_{tt}^{(4)}} \dd t^2
				+ 2 h_{t\mu}^{(3)} \dd t \, \dd x^\mu 
				+ \pac{\delta_{\mu\nu}+h_{\mu\nu}^{(2)}} \dd x^\mu \dd x^\nu
				+ ... \, ,
\end{equation}
where a superscript in brackets indicate the order of smallness of the quantity in terms of $\eps$, and the ellipsis denotes terms of higher order than those given explicitly. For matter fields being modelled as a perfect fluid we end up with
\begin{equation}
\rho = \rho^{(2)} +  \rho^{(4)} + ... \qquad {\rm and} \qquad p = p^{(4)} + ... \, ,
\end{equation}
for the energy density and isotropic pressure, respectively. The orders of accuracy given above are sufficient to model the gravitational physics associated with massive particles to first post-Newtonian order. For the post-Newtonian treatment of light, however, we can neglect $h_{tt}^{(4)}$, $h_{t\mu}^{(3)}$, $\rho^{(4)}$ and $p^{(4)}$. The reader will note that not every order in $\epsilon$ is required in every variable, in order to get a complete description.

The post-Newtonian formalism, as outlined above, is designed for use in the presence of astrophysical bodies like stars and planets. It is, however, unsuitable for direct application to cosmology. This is because on distances comparable to the Hubble radius the coordinate velocity of the cosmological fluid becomes close to the speed of light, and the metric coefficients differ very significantly from the assumed Minkowski background. The perturbative expansion therefore breaks down, and the entire formalism fails. This problem motivates the piecewise construction summarised below, which allows the post-Newtonian approach to be extended to cosmology in a fully self-consistent fashion.

\subsubsection{A bottom-up cosmological model}

\begin{figure}[t!]
\centering
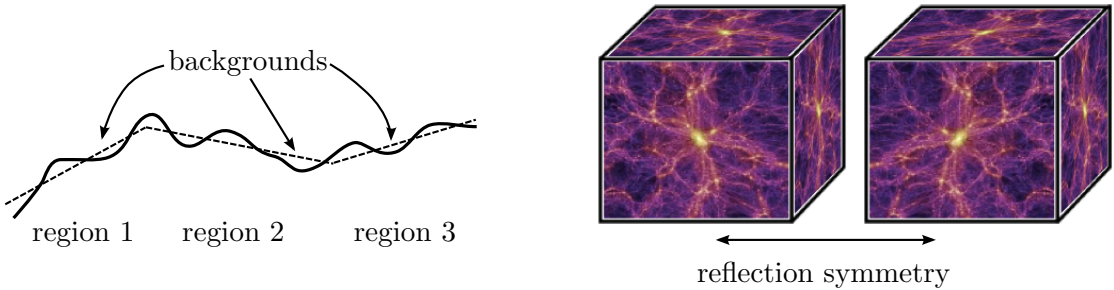
\caption{\textit{Left panel:} A schematic representation of a cosmology constructed piece-wise from regions of post-Newtonian space-time. \textit{Right panel:} Two cells of the periodic cosmological model are symmetric under reflection in their common boundary (picture adapted from Ref.~\cite{2006MNRAS.365...11C}).}
\label{fig:model}
\end{figure}

Consider a region of space-time much, much smaller than the ball bounded by the Hubble sphere. At all points within this region, except in the immediate vicinity of compact objects like neutron stars and black holes, the post-Newtonian formalism should be expected to hold. Now consider a second comparable region, adjacent to the first one. A similar expansion can be performed there, except that the background will not be identical to that of the first region; they will both be regions of Minkowski space, but they will join at an angle to allow for curvature of space-time on cosmological scales (see the left-hand panel of figure~\ref{fig:model}). Thus the whole cosmological space-time can be obtained step-by-step, performing a post-Newtonian expansion in each small region independently, and then by gluing together resulting cells.

Any healthy junction between two space-time regions must fulfil the Darmois-Israel conditions~\cite{Darmois,1966NCimB..44....1I}. These state that, in the absence of surface layers, the intrinsic metric and the extrinsic curvature of the hypersurface that separates any two regions of space-time must be the same whether it is calculated on one side or the other. In order to explicitly ensure these junction conditions, refs.~\cite{2015PhRvD..91j3532S,2016PhRvD..94b3505S} assumed that the Universe could be modelled as being periodic, and that neighbouring cells are symmetric with respect to the boundary that separates them (see the right-hand panel of figure~\ref{fig:model}). In this case, the whole lattice can be generated from a single cell by successive reflections.

Cosmic expansion naturally emerges from the boundary conditions between neighbouring regions, in this model. The large-scale cosmological dynamics can be studied by considering the length of a cell edge, as measured in the rest space of a set of observers that follows the trajectory of the boundary. The analogue of the cosmic scale factor is then given by the ratio~$a(t)\define L(t)/L_0$ between the size of the cell edge at time $t$ and its value today. After extending the post-Newtonian formalism to include radiation and $\Lambda$, the equation that governs the evolution of this scale factor can be shown to be given by~\cite{2016PhRvD..94b3505S}
\begin{equation}
\label{backreaction}
\left( \frac{\dot{a}}{a} \right)^2= \frac{8\pi G}{3} (\rho\e{m}+\rho\e{r}) - \frac{k}{a^2} + \frac{\Lambda}{3} + \mathcal{B},
\end{equation}
where $\rho\e{m}$ is the energy density in pressureless matter, $\rho\e{r}$ is the energy density in radiation, $k$ is spatial curvature, and a dot denotes a derivative with respect to proper time measured by the comoving observers. Equation~(\ref{backreaction}) is essentially the standard Friedmann equation, but with an additional source term~$\mathcal{B}$ on the right-hand side that is of order $\eps^4$. This new term encodes the back-reaction effect of structure within the cell on the large-scale expansion, and for a point-like mass at the centre of the cell takes the simple form
\begin{equation}
\label{B}
\mathcal{B} \approx -\pac{4\pi G \rho\e{m} L(t)}^2 \pa{ 1.50 - 0.80\,\frac{\Omega\e{r}}{\Omega\e{m}} + 1.76\,\frac{\Omega_k}{\Omega\e{m}} }.
\end{equation}
This contribution to the large-scale expansion is purely relativistic, and in a universe dominated by dust and $\Lambda$ can be seen to be principally determined by a term that behaves precisely like an extra radiation component. Such a correction is reminiscent of the outcome from Green and Wald's approach to backreaction~\cite{2011PhRvD..83h4020G,2014CQGra..31w4003G}, and was also found in earlier work on reflection-symmetric subspaces~\cite{2014CQGra..31j5012C}. The $\Omega\e{m}, \Omega\e{r}$, and $\Omega_k$ parameters in eq.~(\ref{B}) denote the cosmological parameters associated with matter, radiation, and curvature respectively.

It is within the context of these lattice models that we will investigate the propagation of light, and the construction of Hubble diagrams. This will be done using opaque compact objects at the centre of each cell, as well as with dispersed transparent matter. We intend these configurations to roughly correspond to the type of matter that exists at the centre of galaxies, and the dark matter haloes within which galaxies are known to exist. The post-Newtonian approach to modelling gravitational fields in these constructions is ideal for addressing old questions about the effect of discretization of matter on light propagation in cosmology~\cite{1964SvA.....8...13Z}, as the matter fields are allowed to be highly non-linear and the intervening regions of vacuum are allowed to be entirely Ricci flat.

\subsubsection{Comparison with other approaches}

The post-Newtonian cosmological models described above can be contrasted with other approaches to inhomogeneous cosmology. Two of the most well-known of these are standard cosmological perturbation theory on an FLRW background~\cite{PeterUzan}, and the Swiss cheese models originally constructed by Einstein and Straus~\cite{1945RvMP...17..120E,1946RvMP...18..148E}. The former of these approaches is of course extremely versatile, but is strictly only valid in the regime where density contrasts are small (of the same order as the expansion parameter). The latter allows for arbitrarily large density contrasts by modelling inhomogeneity as patches of either Schwarzschild~\cite{1945RvMP...17..120E,1946RvMP...18..148E,1969ApJ...155...89K,DR72,1973ApJ...180L..31D,1973PhDT........17D,2013PhRvD..87l3526F,2013PhRvL.111i1302F,2014JCAP...06..054F}, Lema\^{i}tre-Tolman-Bondi~\cite{Marra:2007pm,2008PhRvD..77b3003M,Brouzakis:2007zi,2007JCAP...02..013B,Biswas:2007gi,Vanderveld:2008vi,Valkenburg:2009iw,Clifton:2009nv,Szybka:2010ky,2011JCAP...02..025B,Flanagan:2012yv,2013JCAP...12..051L,2015arXiv150706590L} or Szekeres solutions~\cite{2009GReGr..41.1737B,2010PhRvD..82j3510B,2014PhRvD..90l3536P,2014JCAP...03..040T,2017PhRvD..95f3532K}, but is somewhat less versatile due to its reliance on matter being well-modelled by these exact solutions. A common feature of both of these models, however, is that they can be considered to be top-down constructions, as they assume a global FLRW  background geometry from the outset. This means that the large-scale cosmological behaviour is pre-specified, and cannot be affected by the formation of structure.

In comparison, post-Newtonian cosmological models do not contain any global assumptions at the level of the metric, but do assume that the matter distribution can be modelled as periodic (and hence statistically homogeneous). From this point of view they can be compared with a number of other recent articles on lattice cosmology. The first of these was based on the Lindquist-Wheeler construction~\cite{2009PhRvD..80j3503C,2009JCAP...10..026C,2012PhRvD..85b3502C}, where Schwarzschild regions are glued together using approximate matching conditions in order to form a lattice of point masses~\cite{1957RvMP...29..432L,2015PhRvD..92f3529L}. A second approach is the Bruneton-Larena lattice Universe~\cite{2012CQGra..29o5001B,Bruneton:2012ru}, which is based on perturbatively solving the Einstein field equations after performing a Fourier decomposition, and which results in a cosmological model that can be used for short periods of cosmic time. Further attempts at modelling the Universe as a lattice of point-particles were performed using Regge calculus by Liu and Williams~\cite{Liu:2015bya,Liu:2015gpa,Liu:2015bwa}, and by using geometrostatics and numerical relativity methods by a number of other authors~\cite{Clifton:2012qh,Clifton:2013jpa,Clifton:2016mxx,Durk:2016yja,Clifton:2017hvg,Bentivegna:2012ei,Bentivegna:2013xna,Bentivegna:2013jta,Korzynski:2015isa,Yoo:2012jz,Yoo:2013yea,Yoo:2014boa}. The post-Newtonian method has the advantage of extra versatility, when compared to these other approaches.

As well as the dichotomy between top-down and bottom-up approaches, we could also classify inhomogeneous cosmological models according to whether or not inhomogeneities are `screened' (in the sense that what happens in a given region of a model depends only on its immediate neighbourhood, or if it is affected by all matter that exists in the Universe). By construction, Swiss cheese models and Lindquist-Wheeler lattices belong to the first category, as within a Swiss cheese hole or a Lindquist-Wheeler cell space-time is entirely determined by the closest mass. 
By contrast, most other approaches result in individual inhomogeneities affecting all of the rest of space-time. The comparison of these approaches to inhomogeneous cosmology are summarised in Table~\ref{tab:inhomogeneous_cosmologies}. The reader may note that post-Newtonian cosmology combines the advantages of being background-free (bottom-up), as well as being realistic (not screened) and versatile.

\begin{table}
\renewcommand\arraystretch{1.6}
\renewcommand\tabcolsep{10pt}
\centering
\begin{tabular}{c||c|c}
 & \textbf{Top-Down} & \textbf{Bottom-Up} \\ 
\hline  \hline
\textbf{Screened} & Swiss cheese models & Lindquist-Wheeler lattice \\ 
\hline 
\textbf{Not Screened} & perturbation theory & Bruneton-Larena \& black hole lattices\\
					&	exact solutions		& post-Newtonian cosmology \\ 
\end{tabular} 
\caption{Classification of various approaches to inhomogeneous cosmological modelling. Top-down versus bottom-up refers to whether an FLRW background is assumed or not. Screening refers to whether inhomogeneities can affect the entire space-time, or only their own locale.}
\label{tab:inhomogeneous_cosmologies}
\end{table}

\subsection{Light propagation in curved space-time}
\label{subsec:light_propagation_curved_space-time}

The purpose of this article is to determine the Hubble diagrams that would be constructed by observers in the  post-Newtonian cosmological models, described above. This requires us to understand how light propagates in curved space-times, so that we can determine both distance measures and redshifts. In this section we therefore provide a brief overview of the main results of geometric optics in the presence of gravitation. This will serve to introduce the reader to the subject, as well as to specify the conventions and notation we will use in the rest of the article. For more detailed descriptions see refs.~\cite{1965gere.book..249P,1992grle.book.....S,2004LRR.....7....9P,2015arXiv151103702F}.

\subsubsection{Rays of light}

Light is an electromagnetic wave, and therefore propagates according to Maxwell's equations. Assuming a minimal coupling between the  electromagnetic and gravitational fields, and under the eikonal approximation of geometric optics, it can be shown that rays of light must follow null geodesics. The four-vector~$\vect{k}$ tangent to each ray must therefore satisfy
\begin{equation}
k^a k_a = 0
\end{equation}
and
\begin{equation}\label{eq:geodesic_equation}
\Ddf{k^a}{\lambda} \define k^b \nabla_b k^a = \ddf{k^a}{\lambda} + \Gamma\indices{^a_b_c} k^b k^c = 0 \, ,
\end{equation}
with~$k^a\define \dd x^a/\dd\lambda$ and~${\dd k^a}/{\dd\lambda} \define k^b k\indices{^a_{,b}}$, and where $\lambda$ denotes an affine parameter along the ray of light. If the rays of light are non-rotating then the four-vector components~$k_a$ can also be written as the gradient of the wave's phase, such that~$k_a = \partial_a \phi$.

For an observer following an integral curve of the time-like four-velocity~$\vect{u}$, the wave four-vector~$\vect{k}$ can be split into a temporal part $\omega=-u_a k^a$ and a spatial part $e_a = (g_{ab} +u_a u_b) k^b$. These quantities encapsulate the cyclic frequency and the direction of propagation, respectively. If we now write $\vect{e} = \omega \vect{d}$, then 
\begin{equation}\label{eq:decomposition_k}
\vect{k} = \omega (\vect{u}+\vect{d}) \, ,
\end{equation}
from whence it can be seen that $u_a d^a=0$ and $d^a d_a=1$. In writing eq.~\eqref{eq:decomposition_k} we have chosen $\vect{k}$ to be future oriented, so that $\lambda$ increases to the future.

The redshift involved in most cosmological observations is defined as the fractional difference between the frequency~$\omega$ emitted by a light source and the frequency~$\omega_0$ at which it is observed, such that
\begin{equation}
1+z = \frac{\omega}{\omega_0} \, .
\end{equation}
Now suppose that a collection of sources lie along the light ray, forming a field of four-velocities~$\vect{u}(\lambda)$. In this case the evolution of $z$ with $\lambda$ can be shown to read
\begin{equation}\label{eq:z_lambda}
\ddf{z}{\lambda} = - k^a k^b \nabla_a u_b \define - (1+z)^2 H_{||} \, ,
\end{equation}
where $H_{||}$ is the local rate of expansion of the family of sources in the direction of propagation of the ray of light, and where we have chosen units so that $\omega_0=1$. The function $z=z(\lambda)$ therefore contains information about the expansion of the Universe.

\subsubsection{Narrow beams of light}

To calculate other astronomical observables, such as measures of distance, requires a mathematical description of beams of light. For these purposes we will consider narrow beams of light, which can be modelled as bundles of null geodesics that are infinitesimally separated from each other. The distance between two neighbouring photons within the beam is then described by the separation vector~$\vect{\xi}$, which is defined as being geodesic and orthogonal to~$\vect{k}$. We therefore have $k_a \xi^a=0$ and
\begin{equation}
\label{deveq1}
\Ddf[2]{\xi^a}{\lambda} = R\indices{^a_b_c_d} k^b k^c \xi^d \, ,
\end{equation}
where $R\indices{^a_b_c_d}$ are the components of the Riemann curvature tensor. This last expression is the geodesic deviation equation, which (after projection and integration) allows us to determine the angular diameter distance and luminosity distance along the beam.

In order to study the morphology of the beam it is convenient to project it onto a screen that is orthogonal to its direction of propagation. A set of orthonormal basis vectors that span such a screen can be written as~$(\vect{s}_1,\vect{s}_2)=(\vect{s}_A)_{A=1,2}$, where $u_a s_A^a=d_a s^a_A=0$. The basis vectors on screens at different positions along the beam can then be related by imposing the partial parallel-transportation condition~$(\delta^a_b+u^a u_b - d^a d_b) \Dd s_A^b/\dd\lambda = 0$, which for computational purposes can be more conveniently written as
\begin{equation}\label{eq:Sachs_vector}
\Ddf{s^a_A}{\lambda} = \frac{k^a}{\omega} \Ddf{u_b}{\lambda} s_A^b \, .
\end{equation}
Transporting the basis vectors of the screen-space in this way prevents the beam's morphology from spuriously rotating as one proceeds along its direction of propagation, and results in what is usually referred to as the {\it Sachs basis}.

The components of the separation vector in the Sachs basis can now be written as $\xi_A \define \xi_a s^a_A$. These new objects exist entirely within the plane of the screen, and give us direct information about the separation between photons within the beam (i.e. the separation that an observer would measure if he or she replaced the screen with a photographic plate). Using eq. (\ref{deveq1}), the projected separation vectors can be seen to obey the following evolution equation:
\begin{equation}
\label{deveq2}
\ddf[2]{\xi_A}{\lambda} = \tidal_{AB} \xi_B \, ,
\end{equation}
where $\tidal_{AB} \define R_{abcd} s^a_A k^b k^c s_B^d$ is referred to as the {\it optical tidal matrix}, and the whole equation is known as the {\it vector Sachs equation}. The reader may note that the position of the indices~$A,B,\ldots$ does not matter, as they are raised and lowered by $\delta_{AB}$. The optical tidal matrix can be decomposed into a scalar part and a trace-free part as
\begin{equation}
\label{tidal}
\vect{\tidal} =
\begin{pmatrix}
\Ricfoc & 0\\
0 & \Ricfoc
\end{pmatrix}
+
\begin{pmatrix}
-\Weylfoc_1 & \Weylfoc_2 \\
\Weylfoc_2 & \Weylfoc_1
\end{pmatrix}.
\end{equation}
The scalar part depends only on the Ricci curvature~$R_{ab}$ of the space-time, with $\Ricfoc=-(1/2) R_{ab} k^a k^b$, and causes the beam of light to be focussed without distortion. Meanwhile, the trace-free part depends only on the Weyl curvature~$C_{abcd}$ of the space-time, with $\Weylfoc_1\define C_{abcd} s_1^a k^b k^c s_1^d=C_{abcd} s_2^a k^b k^c s_2^d$, $\Weylfoc_2\define C_{abcd} s_1^a k^b k^c s_2^d$, and causes distortion of the beam (by tidal gravitational forces).

Because the vector Sachs equation~\eqref{deveq2} is linear in $\xi_A$, any solution must be linearly related to its initial conditions at $\lambda_0$. More precisely, there must exist a $4\times 4$ Wronski matrix~$\vect{\wronski}$ such that
\begin{equation}\label{eq:Wronski}
\begin{pmatrix}
\xi_1 \\
\xi_2 \\
\dot{\xi}_1 \\
\dot{\xi}_2
\end{pmatrix}
(\lambda)
=
\vect{\wronski}(\lambda\leftarrow\lambda_0)
\begin{pmatrix}
\xi_1 \\
\xi_2 \\
\dot{\xi}_1 \\
\dot{\xi}_2
\end{pmatrix}
(\lambda_0) \, ,
\end{equation}
where an over-dot denotes $\dd/\dd\lambda$, and where~$\vect{\wronski}(\lambda_0\leftarrow\lambda_0)=\vect{1}_4$. From eq.~(\ref{deveq2}), we can then write
\begin{equation}
\label{deveq3}
\ddf{\vect{\wronski}}{\lambda}
=
\begin{pmatrix}
\vect{0}_2 & \vect{1}_2 \\
\vect{\tidal} & \vect{0}_2
\end{pmatrix}
\vect{\wronski},
\end{equation}
where $\vect{0}_n$ and $\vect{1}_n$ denote the $n\times n$ zero and unity matrices, respectively. This reduces the problem of finding $\xi_A$ along the beam to solving a set of first-order ordinary differential equations. It also allows us to use the following matrix multiplication rule:
\begin{equation}\label{eq:Wronski_multiplication}
\vect{\wronski}(\lambda_2\leftarrow\lambda_0) 
= \vect{\wronski}(\lambda_2\leftarrow\lambda_1) \vect{\wronski}(\lambda_1\leftarrow\lambda_0) \, ,
\end{equation}
which is extremely advantageous when considering a cosmological model that is constructed in a piecewise fashion, as we are doing here. We will therefore use eq.~(\ref{deveq3}) as the final form of the geodesic deviation equation when we numerically solve these equations in section~\ref{sec:results}. 

Although we will integrate eq.~(\ref{deveq3}) to find $\vect{\wronski}$ at all values of $\lambda$, the most interesting parts of this matrix are the $2\times 2$ cells in the top-right corner. These are collectively referred to as the {\it Jacobi matrix}, $\vect{\jacobi}$, and are sufficient to determine the angular diameter distances of astrophysical objects that lie within the beam, as we will now discuss.

\subsubsection{Distance measures}

The first step in calculating the angular diameter distance of a source is recognising that the initial condition at the point of observation should be taken to be~$\xi_A(\lambda_0)=0$, as this is the point at which the beam converges. The projection of the separation vector~$\xi_A(\lambda)$ is then linearly related to $\dot{\xi}_A(\lambda_0)$ only, and must therefore be given by the Jacobi matrix~$\vect{\jacobi}=(\jacobi_{AB})_{A,B=1,2}$. This gives
\begin{equation}\label{eq:Jacobi}
\xi_A(\lambda) = - \jacobi_{AB} \theta_B \, ,
\end{equation}
where $\theta_A=-\dot{\xi}_A(\lambda_0)$ is the angular separation between the two light rays separated by~$\vect{\xi}$, as measured at the point of observation. In other words, the matrix $-\vect{\jacobi}$ is the map from observed angular separations to spatial separations in the screen space at $\lambda$. This is exactly what is required to define the angular diameter distance.

The definition of the angular diameter distance to a source with cross-sectional area $A_S$, and that subtends the angle $\Omega_0$ on the observer's sky, is
\begin{equation}
D\e{A} \define \sqrt{\frac{A_S}{\Omega_0}} \, .
\end{equation}
This quantity is defined in analogy to the way that one would infer distance in a flat space-time, if the same source subtended the same angle. Using eq.~\eqref{eq:Jacobi}, and recognising that the area of a parallelogram is given by the determinant of the matrix formed from the vectors that define it, one can directly deduce that
\begin{equation}
D\e{A} = \sqrt{\det \vect{\jacobi}} \, .
\end{equation}
Once the Wronskian matrix $\vect{\wronski}$ has been obtained, the angular diameter distance at all points along the beam can therefore be readily deduced by simple algebraic operations. This gives $D\e{A}$ as a function of $\lambda$, and can be used to obtain $D\e{A}$ as a function of redshift by using the solution of eq.~(\ref{eq:z_lambda}). 

From knowledge of the angular diameter distance, it is relatively straightforward to obtain an expression for the luminosity distance. This latter measure is defined as
\begin{equation}
D\e{L}\define \sqrt{\frac{L}{4\pi I}} \, ,
\end{equation}
which would be the distance that one would infer in a flat space for a sources that has luminosity $L$, and is measured to have intensity $I$. In this case the beam of light must be taken to be focussed at the emitting source, such that $\xi_A(\lambda_S)=0$. If the cross-sectional area of this beam is $A_0$ at the observer, and the angle it subtends at the source is $\Omega_S$, then by photon conservation the luminosity distance is simply given by
\begin{equation}
\label{receq}
D\e{L}= (1+z)\sqrt{\frac{A_0}{\Omega_S}} = (1+z)^2 D\e{A}\, .
\end{equation}
The last of these equalities comes from Etherington's reciprocity theorem, which states that $A_0 \Omega_0 = (1+z)^2 A_S \Omega_S$ in any space-time \cite{1933PMag...15..761E}. The remarkable nature of this theorem means that $D\e{L}(z)$ can now be obtained without integrating the light beam forward from the source, if one already has knowledge of the beam that is focussed at the observer.

%

\section{Method}
\label{sec:method}

Let us now turn to the detailed implementation of light propagation in post-Newtonian cosmology using ray tracing techniques. In order to capture the leading-order post-Newtonian effects on the null geodesics that constitute the paths followed by individual rays of light we require the metric to be specified to order~$\eps^2$:
\begin{equation}\label{eq:metric}
\dd s^2 = -(1-2\Phi) \dd t^2  + (1+2\Psi) \delta_{\mu\nu} \dd x^\mu \dd x^\nu + \mathcal{O}(\eps^3) \, .
\end{equation}
The scalar gravitational potentials $\Phi$ and $\Psi$ are both objects of order~$\eps^2$ in the post-Newtonian expansion of the metric, and it can be noted that at this order there can exist no vector or tensor perturbations. The required geometric quantities associated with this metric are given in appendix~\ref{app:geometry}, up to order~$\eps^2$. While the form of these scalar potentials is similar to those used in standard cosmological perturbation theory, their precise meaning is different.  In particular, the functions~$\Phi$ and~$\Psi$ in eq.~(\ref{eq:metric}) contain information about the global expansion, as well as the gravitational fields of nearby objects.

\subsection{A universe in a cell}
\label{subsec:cell_properties}

\begin{figure}[t!]
\centering
\def\svgwidth{10cm}
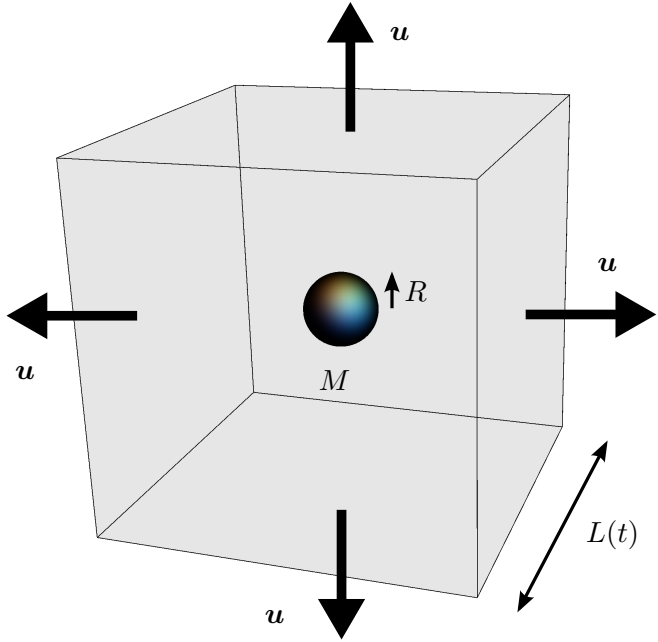
\caption{An illustration of the cubic lattice cell we are considering, with a spherical ball of matter with mass $M$ and radius~$R$. The size of the cell, $L(t)$, evolves with time, and the four-velocity of the faces serve to define a four-velocity field~$\vect{u}$ that can be continued into the interior of the cell.}
\label{fig:cell}
\end{figure}

As the space-time we are considering is periodic, we only need to describe the properties of a single cell in order to get the geometry of the entire universe. The tessellation of space that we wish to consider here is based on a cubic cell, which we use to tile a flat three-dimensional reference space. In this case, the volume of the lattice cell can be taken to be~$V_0 \simeq L_0^3=1\U{Mpc}^3$ today, where $L_0$ is the proper length of one of its edges at the present time. The condition that the lattice is constructed in a flat space means that the total rest mass within each cell must be exactly specified by~\cite{2015PhRvD..91j3532S,2016PhRvD..94b3505S}
\begin{equation}
M = \rho_0 L_0^3 = \frac{3 H_0^2 \Omega\e{m} L_0^3}{8\pi}
\approx 4.5 \times 10^{10} M_\odot \, ,
\end{equation}
where $H_0$ is the current value of the Hubble rate, $\rho_0$ the current mean density of matter (both dark and baryonic) in the Universe, and~$\Omega\e{m}=0.3$ its ratio with the critical density. In what follows we will assume this mass to be contained within a spherically symmetric static body at the centre of the cell, with radius~$R\ll L_0$. This setup is illustrated in figure~\ref{fig:cell}. One may note that other tessellations exist, in three-dimensional spaces of positive and negative curvature.

We intend the matter within each cell to represent either a galaxy, or a galactic dark matter halo. In the rest of this article we will therefore consider two different simulations:
\begin{description}
\item[(i) Galaxy simulations:] In this case the massive body at the centre of the cell models the compact bulge of a spiral galaxy. We take this object to have radius~$R=3\U{kpc}$, and assume it to be \emph{opaque}. This last condition is considered to be largely true in the real Universe, where the density of luminous matter and gas in a bulge makes it almost impossible to view any objects that lie behind it. In practice, opacity is ensured in our models by artificially removing from the simulation any rays that enters into the region $r<3\U{kpc}$. This setup is comparable to the Einstein-Straus Swiss cheese model~\cite{1945RvMP...17..120E,1946RvMP...18..148E}, and the black-hole lattice considered in ref.~\cite{Bentivegna:2016fls}.
\item[(ii) Halo simulations:] Here the central body represents a diffuse dark matter halo. We take this object to have radius~$R=30\U{kpc}$, and to be completely \emph{transparent}. We assume for simplicity that the halo has uniform density~$\rho\e{H}=3M/4\pi R^3$. This is far from being fully realistic, but is expected to be sufficient to address the questions we pose in this article. More realistic density profiles will be implemented in the future. This setup is comparable to those that are often considered in the context of standard cosmological perturbation theory, as well as the LTB and Szekeres Swiss cheese models.
\end{description}

\noindent In both cases (i) and (ii) the gravitational potentials can be shown to be given by~\cite{2015PhRvD..91j3532S,2016PhRvD..94b3505S}
\begin{align}\label{eq:potential}
\Phi(t,\vect{x}) &= \Phi_0(\vect{x}) + \pa{\sum_{\vect{p}\in\mathbb{Z}^3_*} \frac{M}{|\vect{x}-L(t)\vect{p}|} - \frac{M}{|L(t)\vect{p}|} } + \frac{\Lambda}{6} \, , 
\end{align}
and $\Psi(t,\vect{x}) = \Phi(t,\vect{x}) - {\Lambda}/{4}$, where the origin of the spatial coordinate system lies at the centre of the cell, and where $\Phi_0(\vect{x})$ is given by
\begin{equation}
\Phi_0(\vect{x})
=
\begin{cases}
-\dfrac{M}{2 R^3} (\vect{x}^2-3 R^2) & \text{if }|\vect{x}|\leq R \\[3mm]
\dfrac{M}{|\vect{x}|} & \text{if } |\vect{x}|\geq R \, .
\end{cases}
\end{equation}
Note that the case $r\leq R$ is relevant for the halo simulations only, as in the galaxy simulations light is not allowed to enter this region.

To the same order of accuracy, it can be shown that the global expansion in both cases (i) and (ii) is given by the following emergent Friedmann equations:
\begin{equation}\label{eq:Fried1}
\pa{\frac{\dot{a}}{a}}^2 = \frac{8 \pi G\rho}{3} + \frac{\Lambda}{3} + \mathcal{O}(\eps^4) \qquad {\rm and} \qquad
\frac{\ddot{a}}{a} = -\frac{4 \pi G\rho}{3} + \frac{\Lambda}{3} + \mathcal{O}(\eps^4) \, ,
\end{equation}
where $\rho=M/L^3$ is the averaged energy density of matter in each cell (to order~$\eps^2$), and where $\Lambda$ is the cosmological constant.  At this level of accuracy, these two equations are identical to the standard Friedmann equations of homogeneous and isotropic cosmological models. They are, however, derived from the Israel junction conditions applied at the boundaries between neighbouring cells, rather than by assuming that averaged energy densities and geometries can be substituted into Einstein's equations directly. This means that they are valid for arbitrary distributions of matter within each cell, including the types of highly non-linear density contrasts required to describe galaxies and clusters of galaxies. Once these equations have been solved, the length of the cell edge is given by $L(t)=a(t) L_0$, where $a(t_0)=1$.

The final piece of information required to compute observables such as redshift~$z$ and angular diameter distance~$D\e{A}$ is a four-velocity field~$\vect{u}$ that covers the whole of the space-time (or, equivalently, the whole of a cell). Once specified, $\vect{u}$ can be taken to define the rest frame of a hypothetical light source at any point in the space-time. We require observers who follow the integral curves of this field to be comoving with the boundary when they are coincident with it, and to be at rest at the centre of the cell if they are located at that position. In other words, we choose our hypothetical sources on the boundary to be comoving with the boundary. A vector field that obeys these conditions, and that has been continued into the interior of the cell, is given by~\cite{2015PhRvD..91j3532S}
\begin{equation}\label{eq:four_velocity_comoving}
\vect{u} = 
\pa{ 1 + \Phi + \frac{1}{2} H^2 \vect{x}^2 } \vect{\partial}_t
+ H x^\mu \vect{\partial}_\mu
+ \mathcal{O}(\eps^3) \, ,
\end{equation}
where $H\define \dot{a}/a$. The observers who follow the integral curves of this field are analogous to the fundamental (comoving) observers used in standard FLRW cosmology. We use them to define the frequency of light, the Sachs basis, and every other frame-dependent variable.

\subsection{Initial conditions}
\label{subsec:initial_conditions}

The ray tracing procedure outlined in section~\ref{subsec:light_propagation_curved_space-time} requires appropriate initial conditions to be set. Specifically, we need to specify the position of the observer in both space and time, as well as the direction on the observer's sky in which the beam of light will be propagated. Once this information has been given, the tangent vector to the light ray, the Sachs basis vectors, and the Wronski matrix will be found by integrating eqs.~\eqref{eq:geodesic_equation}, \eqref{eq:Sachs_vector}, and \eqref{deveq3} backwards in time, from the observation event $O$ to its source at affine distance $\lambda$. 

We choose the coordinates of $O$ by first performing a coordinate transformation, as follows:
\begin{equation}
\begin{aligned}
t_0 &= \hat{t}_0 + \frac{H_0 \hat{r}_0^2}{2} + \mathcal{O}(\eps^3)\\
x^\mu_0 &= \hat{x}^\mu_0 \pac{ 1 - \pa{\frac{H_0 \hat{r}_0}{2}}^2 } + \mathcal{O}(\eps^4) \, ,
\end{aligned}
\end{equation}
where hatted coordinates~$\hat{x}^a$ are the analogue of comoving synchronous coordinates in the post-Newtonian cosmological framework~\cite{2015PhRvD..91j3532S}, and where $\hat{r}_0 \define \sqrt{\hat{x}_0^2+\hat{y}_0^2+\hat{z}_0^2}$. We want our observer to remain at fixed position with respect to these coordinates, so they are as similar as possible to the comoving observers used in FLRW cosmology. We then make the further choice that the time of observation is at
\begin{equation}
\hat{t}_0 = \frac{2}{3H_0\sqrt{\Omega_\Lambda}} \, \mathrm{arcsinh}  \sqrt{\frac{\Omega_\Lambda}{\Omega\e{m}}}  \, ,
\end{equation}
which is the time at which an observer in this model exists in order to measure $H=H_0$, for any given values of $\Omega_{\Lambda}$ and $\Omega\e{m}$. The remaining comoving spatial coordinates are chosen to be $\{\hat{x}_0 , \hat{y}_0  , \hat{z}_0 \} = \{ -0.4 L_0 , 0.1 L_0 , 0  \}$. This places the observer in the bulk of the cell, far from the central mass, the cell edge, and all axes of discrete rotational symmetry~\cite{2014CQGra..31j5012C}.

If the observer at $O$ is comoving, then his or her four-velocity~$\vect{u}_0$ is given by eq.~\eqref{eq:four_velocity_comoving} at $O$. The rest space of such an observer is then spanned by a triad~$(\vect{e}_\alpha)_{\alpha=1,2,3}$ with components
\begin{align}
e_\alpha^0 &= H_0 x_0^\alpha + \mathcal{O}(\eps^3) \, , \\
e_\alpha^\mu &= (1-\Psi_0) \delta^{\mu}_{\alpha} + \frac{1}{2}H_0^2 x_0^\mu x_0^\alpha + \mathcal{O}(\eps^3) \, ,
\end{align}
so that $(\vect{u}_0,\vect{e}_\alpha)$ forms an orthonormal basis at $O$. Our choice of units of time is now such that the observed frequency of light at $O$ is given by~$\omega_0\define -(u^a k_a)_0=1$, and our specification of a spatial triad means the direction of incoming photons on the observer's celestial sphere be written as
\begin{equation}
\vect{d}_0 = d_0^\alpha \vect{e}_\alpha 
= -\sin\theta \cos\ph\,\vect{e}_1 - \sin\theta \sin\ph \vect{e}_2 - \cos\theta \vect{e}_3 \, ,
\end{equation}
where $\theta$ and $\phi$ are standard spherical coordinates. Once the pair of coordinates~$(\theta,\ph)$ have been chosen, the initial conditions for the four-vector tangent to the rays of light can then be seen to become
\begin{align}
k^t_0 &=  1 +  \Phi_0 + \frac{1}{2} H_0^2 r_0^2 + H_0 d_0^\alpha x_0^\alpha \, ,  \\
k^\mu_0 &= H_0 x_0^\mu + (1-\Psi_0) \delta^\mu_\alpha d_0^\alpha + \frac{1}{2} H_0^2 d_0^\alpha x^\alpha_0 x_0^\mu \, .
\end{align}
For each light ray we randomly pick an observation direction, given by~$(\theta,\ph)$, in such a way that the observer's celestial sphere is homogeneously covered; the associated probability density function thus reads $p(\theta,\ph) = {\sin\theta}/{4\pi}$ if $\theta\in[0,\pi]$ and $\ph\in[0,2\pi)$, and zero otherwise. This fully specifies all of the initial conditions for our ray tracing experiment.

\subsection{Reflection of light at the cell's boundary}
\label{subsec:reflection}

\begin{figure}[h!]
\centering
\def\svgwidth{6cm}
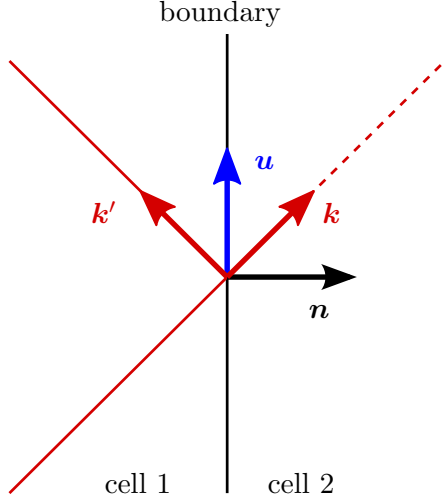
\vspace{0.2cm}
\caption{A schematic illustration of the reflection of light at the boundary, with time increasing in the vertical direction and spatial position varying along the horizontal. The vector~$\vect{n}$ denotes the normal to the boundary, and~$\vect{u}$ and~$\vect{k}$ represent the four-velocities of the boundary (black line) and the ray of light (red line), respectively. The reflection symmetry of the model implies that continuously crossing a boundary from cell 1 to cell 2 is entirely equivalent to reflecting all quantities at the boundary.}
\label{fig:reflection}
\end{figure}

The periodic nature of our lattice means that instead of propagating light rays between neighbouring cells, which is the physical situation we wish to investigate, we can simply reflect our light rays off the cell boundaries and continue propagating it within the same cell. This will produce exactly the same result as propagating the light between cells, as we have reflection symmetry about each of our boundaries (see figure~\ref{fig:reflection}). It is also a more economical way of modelling an infinitely extended universe.

The reflection relations required at our cell boundaries can be derived using the fact that all optical quantities (wave four-vector, Sachs basis, Jacobi matrix, etc.) must be continuous. This fact is ensured to be true at our cell boundaries due to the satisfaction of the Israel junction conditions between neighbouring cells, and the fact that~$\vect{u}$ has been taken to be comoving with the boundary. Thus, for any four-vector~$\vect{v}$ attached to the light beam, its reflected counterpart~$\vect{v}'$ must read
\begin{equation}\label{eq:reflection_relation}
\vect{v}'=\vect{v} - 2\vect{g}(\vect{v},\vect{n})\,\vect{n} \, ,
\end{equation}
where $\vect{n}$ is the outward-pointing unit normal four-vector to the boundary, which for boundaries at coordinate positions~$x^\alpha = \pm L(t)/2$ is given by
\begin{align}
\vect{n} = n^a \vect{\partial}_a
&= +\frac{H L}{2} \vect{\partial}_t \pm \pac{ 1-\Psi + \frac{H^2L^2}{8} } \vect{\partial}_\alpha + \mathcal{O}(\eps^3) \, .
\end{align}
The reflection equation~(\ref{eq:reflection_relation}) applies in particular to the wave four-vector~$\vect{k}$, Sachs basis vectors~$\vect{s}_A$, and separation four-vector~$\vect{\xi}$. 
%
The reader may note, however, that by construction we have~$\vect{u}'=\vect{u}$, and hence $\omega'=\omega$. A further consequence of eq.~\eqref{eq:reflection_relation}, applied to~$\vect{s}_A$ and ~$\vect{\xi}$ is that the screen components of $\vect{\xi}$ are unchanged by the reflection:
\begin{equation}
\xi'_A \define \vect{g}(\vect{\xi}',\vect{s}'_A) 
= \vect{g}(\vect{\xi},\vect{s}_A) \define \xi_A \, .
\end{equation}
This means the Jacobi matrix, the Wronski matrix, and all distance measures must be continuous at reflection, as required. It also means we can apply the Wronski multiplication rule~\eqref{eq:Wronski_multiplication} as if there were no reflection at all.

\subsection{The ray-tracing code}
\label{subsec:code}

\begin{figure}[h!]
\hspace{-1cm}\includegraphics[width=17cm]{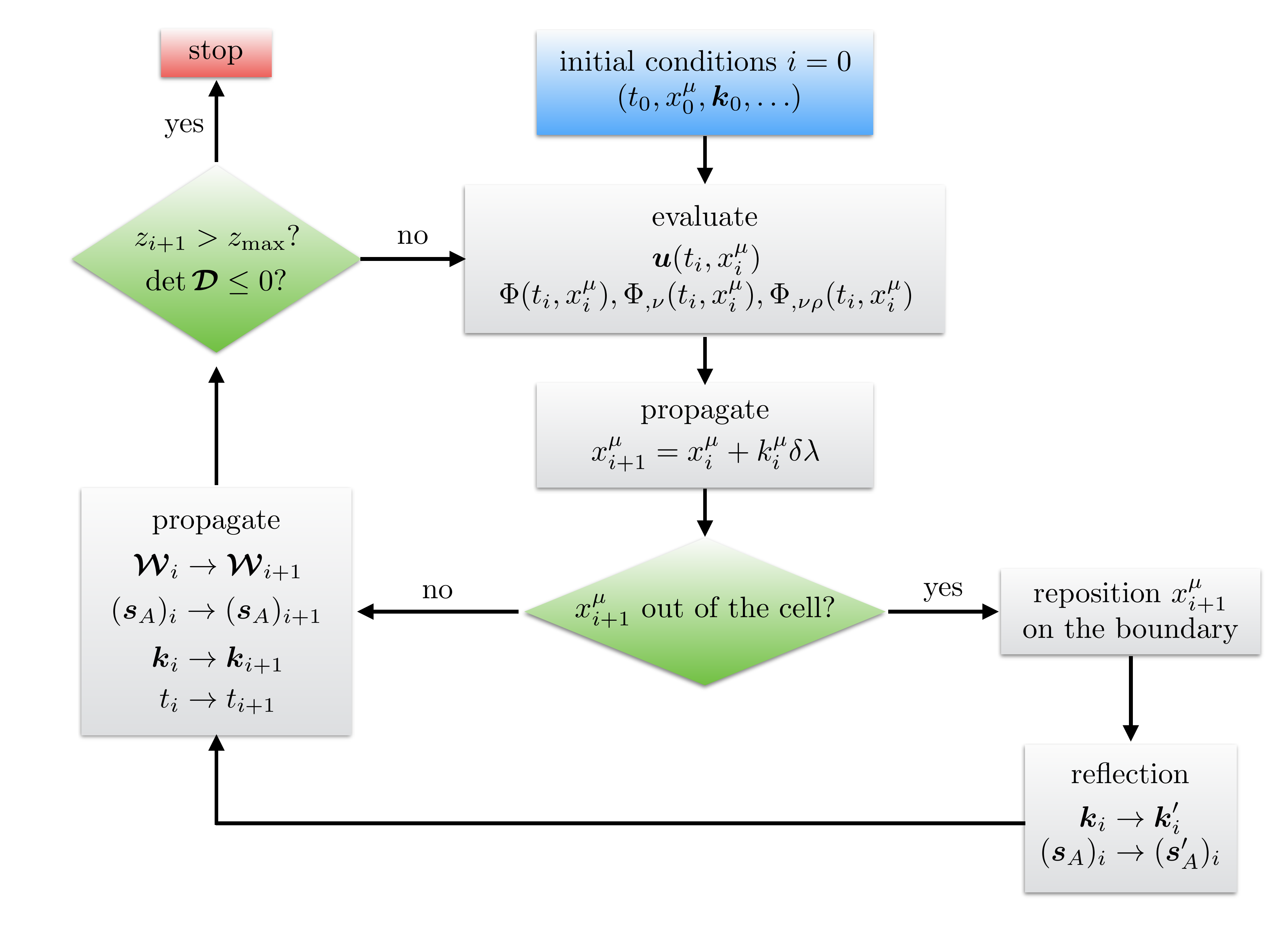}
\caption{A pictorial representation of our ray-tracing code.}
\label{fig:code}
\end{figure}

To implement the integration of the optical equations, as outlined in section~\ref{subsec:light_propagation_curved_space-time}, we developed a ray-tracing code in C. Initial conditions for the numerical integrations we perform are set as described in section~\ref{subsec:initial_conditions}, and the reflection operations at cell boundaries are performed as outlined in section~\ref{subsec:reflection}. The pipeline that implements this set of operations is sketched in figure~\ref{fig:code}.

The code is iterated to look-back time~$t_0-t$ (in the coordinates of eq.~(\ref{eq:metric})), with an evolving step~$\Delta t_i \define t_{i}-t_{i+1}= a(t_i) 1.67\times 10^4\U{yr}$. This choice is made so that the ratio between the time step and the cell size~$L(t)$ remains essentially constant, at $\Delta t/L(t)=5\times 10^{-3}$. This ensures that the accuracy of the code is stable over time. At each time step, we evaluate the gravitational potential~$\Phi$ and its derivatives using eq.~\eqref{eq:potential}. The sum over $\vect{p}$ in this equation is truncated to $\vect{p}\in\{-5,5\}^3\setminus\vect{0}$, so that only the $11^3=1,331$ closest masses are taken into account. The error that is made on the value of the potential due to this truncation is at most
\begin{equation}
\abs{\frac{\Phi\e{exact}-\Phi\e{trunc}}{\Phi\e{exact}}} \sim 5\times 10^{-3} \, ,
\end{equation}
which essentially occurs in the region of space closest to the boundaries of the cell (where the potential is smallest). We expect this to have a negligible impact on our final results. The maximum value of the PN expansion parameter~$\eps$ is evaluated at the edge of the cell to be~$\eps\sim H_0 L_0\sim 2\times 10^{-4}$; in terms of the gravitational potential, it is maximum at the surface of the mass clumps, with~$\eps\sim \sqrt{GM/R}\sim 10^{-3}$ in the galaxy simulation.

Meanwhile, the differential equations of light propagation~\eqref{eq:geodesic_equation}, \eqref{eq:Sachs_vector}, and \eqref{deveq3} are solved using a simple Euler integration with the above time step. The global numerical error on the redshift along an individual line of sight is estimated to be $N\times (\Delta t / L\e{H})^2$, where $N$ is the number of steps, and $L\e{H}$ is the Hubble radius (which is the typical distance over which $\vect{u}$ varies appreciably). As the integration is performed over cosmological distances we have $N\sim L\e{H}/\Delta t$, so that
\begin{equation}
\text{global error on $z$} \sim H_0 \Delta t \sim 10^{-6} \, .
\end{equation}
We have tested our code against known exact expressions in de Sitter space-time, with the results detailed in appendix~\ref{app:tests}. We find that the code exhibits an accuracy for $z(\lambda)$ in agreement with the above prediction. Estimating the global error on the Euler integration of the Jacobi matrix is more subtle. On the one hand, the local error is larger because the optical tidal matrix varies over distances comparable to $L\ll L\e{H}$. On the other hand, its sign flips during the propagation through a cell---curvature increases while the photon is approaching the central clump, and then decreases while it moves away---so that local errors do not combine cumulatively. A conservative estimate can be obtained by considering a random local error with standard deviation $(\Delta t/L)^2$, which along an individual line of sight yields
\begin{equation}
\text{global error on $\vect{\jacobi}$} 
\sim \sqrt{N} \times \pa{\frac{\Delta t}{L}}^2
\sim 10^{-3} \, .
\end{equation}
Numerical tests of the convergence of our numerical integrations indicate a slightly smaller error, below one part in a thousand. This error cannot be tested with the de Sitter case presented in appendix~\ref{app:tests}, however, as the optical tidal matrix is exactly zero in that case.

We will propagate a given light beam within our lattice cell until it reaches a boundary, at which point it will be reflected according to the rules described in section~\ref{subsec:reflection}. Due to the discrete nature of the time steps in the numerical integration, the actual intersection between the photon's path and the world-sheet of the boundary generically occurs between two steps $i$ and $i+1$, so that $x^\mu_{i+1}$ lies outside of the cell. We correct for this by determining the actual point of intersection, and propagating the ray of light back to this point. Observables such as the redshift~$z$, the Jacobi matrix~$\vect{\jacobi}$, and the associated angular distance~$D\e{A}$ are computed at each time step, and saved in an output file at every $z=0.1 n$, with $n\in\mathbb{N}$. 

Finally, the code stops along each ray of light if any one of three cases occurs:
\begin{enumerate}
\item The maximum redshift~$z\e{max}=1.5$ is reached;
\item A caustic occurs, at which point $\det\vect{\jacobi}\leq 0$, and our approach breaks down;
\item The ray of light enters an opaque clump of matter (in the galaxy simulations only).
\end{enumerate}

\section{Results}
\label{sec:results}

\begin{figure}[t!]
\centering
\includegraphics[width=0.4\columnwidth]{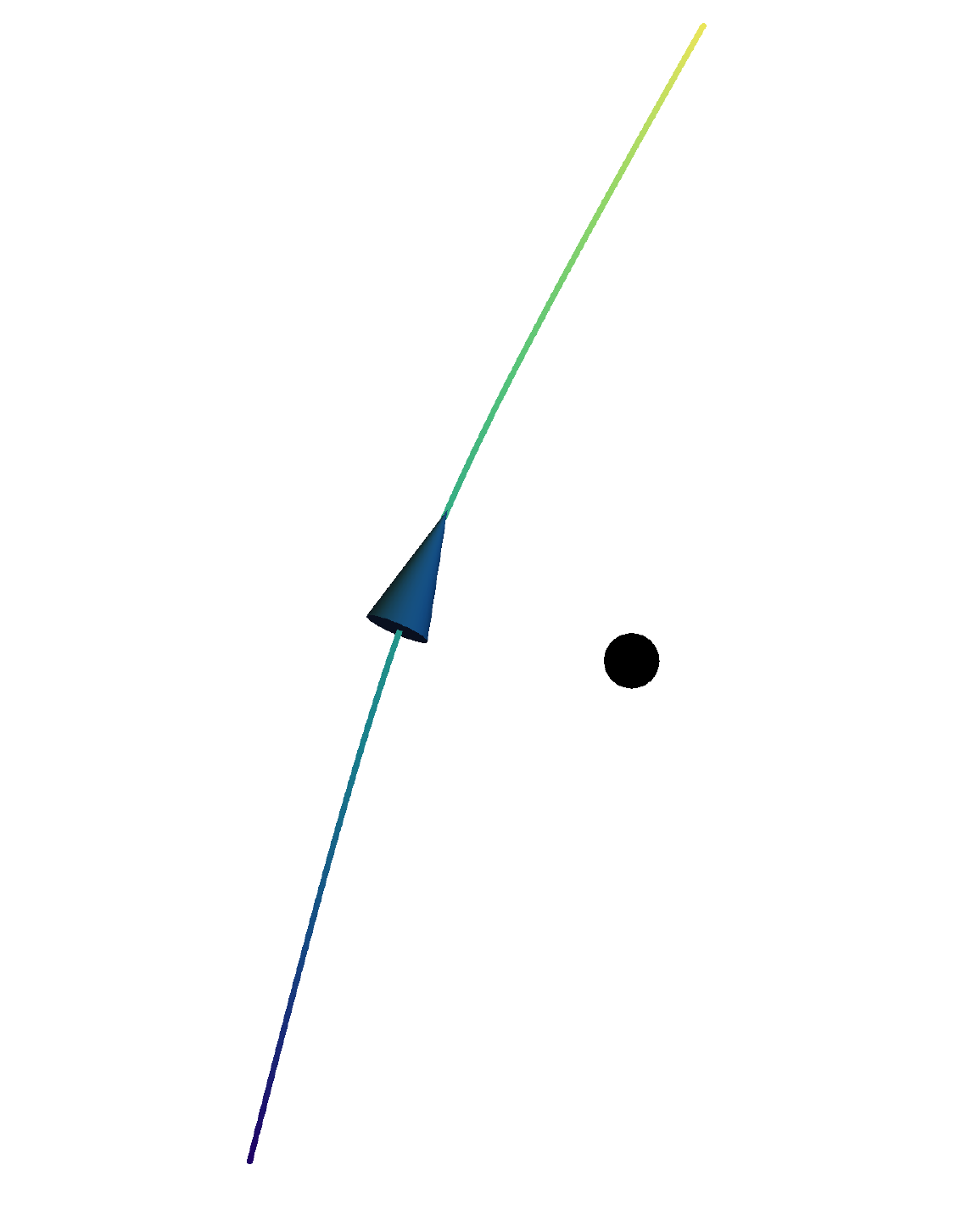}
\hfill
\includegraphics[width=0.55\columnwidth]{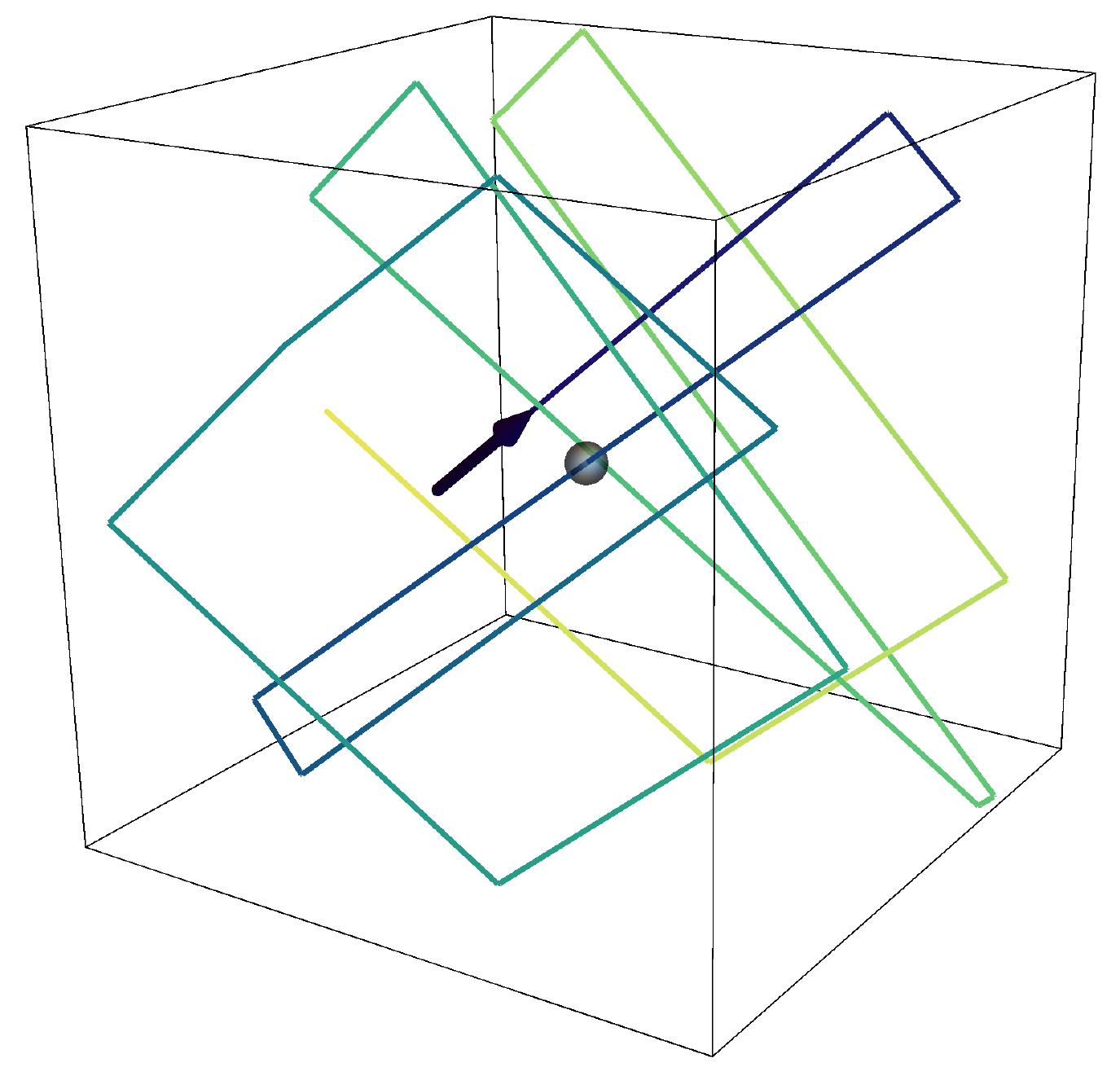}
\caption{\textit{Left panel:} Deflection of light passing very close to a compact object. The black sphere indicates the Schwarzschild radius of the mass~$r\e{S}=2 GM$, and the impact parameter of the ray of light is $10\,r\e{S}$. \textit{Right panel:} The first 20 reflections of a typical realisation of a ray of light in the halo model. The central sphere indicates a halo with radius~$R=30\U{kpc}$, while the box size is $L_0=1\U{Mpc}$.}
\label{fig:illustrations}
\end{figure}

Let us now present the results produced by our ray-tracing code in post-Newtonian cosmologies, before going on to compare these results with existing analytical and numerical approaches in section \ref{sec:discussion}. Our code is constructed so that it can build up statistics associated with physical observables, such as redshift and angular diameter distance, by integrating along large numbers of individual lines-of-sight. Two examples of these individual paths are shown, for illustrative purposes, in figure~\ref{fig:illustrations}. The left panel of this figure shows the deflection of light that occurs close to a compact object, while the right panel shows the first 20 reflections of a beam of light that stays far from the central mass. In both of these diagrams time flows from light to dark colours, while the arrows indicate the direction of numerical integration. Production of the first of these images required us to dramatically reduce the time step of the numerical integrations, in order to resolve the gravitational potential in the vicinity of the compact object\footnote{Note that any light ray passing this close to a mass point in our galaxy simulations would be excluded, due to our selection rules.}.

Let us now address the central question of the statistics of observables that are calculated from considering many such beams of light, and how these relate to the predictions of the homogeneous and isotropic FLRW models. We will do this by first calculating the redshift as a function of affine distance in section~\ref{subsubsec:z}, before moving on to consider the angular distance-affine parameter relation in section~\ref{subsubsec:D_A}. With both of these data sets in hand, we will then combine them to construct Hubble diagrams in section~\ref{sec:hubble}. This last quantity is the direct observable, and is what astronomers within our space-time should be expected to measure. In practice, for each simulation (galaxy and halo) we shoot $N_{\rm L}=10^5$ light beam in random directions on the observer's celestial sphere. The statistical average of an observable~$Q$ is then defined by
\begin{equation}
\ev{Q}_\Omega \define \frac{1}{N_{\rm L}} \sum_{b=1}^{N_{\rm L}} Q_b \, ,
\end{equation}
which can be seen to correspond to a directional average, as the random set of beams evenly covers the sky.
 
\subsection{Redshift}
\label{subsubsec:z}

\begin{figure}[t!]
\centering
\includegraphics[width=0.49\columnwidth]{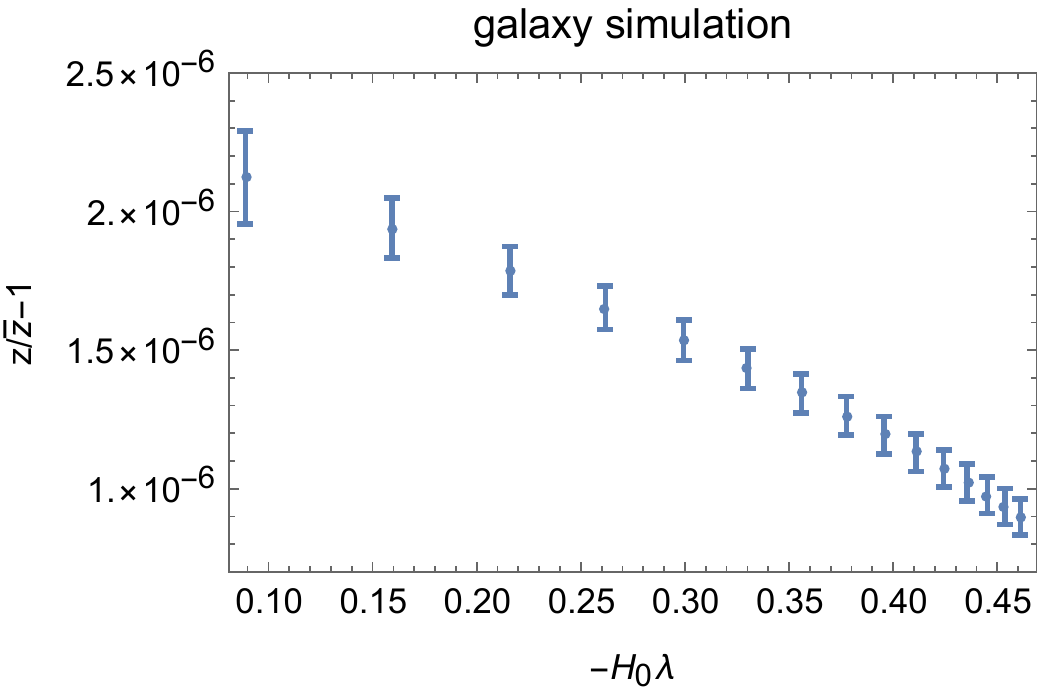}
\hfill
\includegraphics[width=0.49\columnwidth]{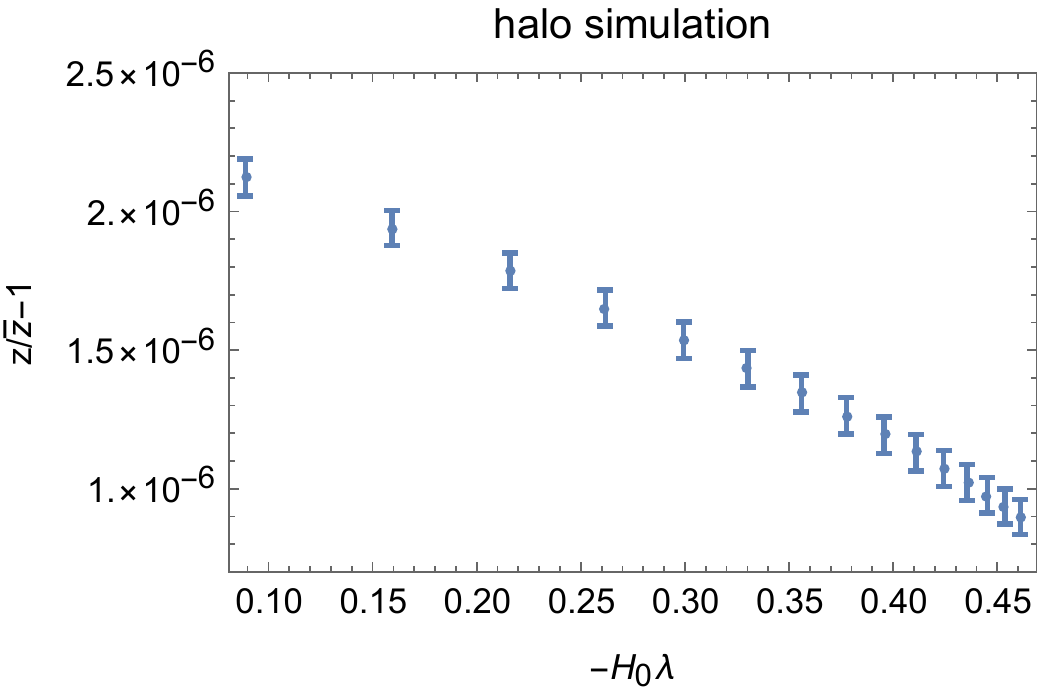}
\caption{Comparison of the redshift-affine parameter relation~$z(\lambda)$ obtained from numerical integration of rays of light in our post-Newtonian simulations, and the corresponding function~$\bar{z}(\lambda)$ in an FLRW model. Dots indicate the statistical mean~$\ev{z}_\Omega/\bar{z}-1$ over $10^5$ light beams shot in random directions from a single location. The error bars indicate the standard deviation~$\sigma_z/\bar{z}$ of $z$ within this data set. The errors associated with numerical precision and truncation of the post-Newtonian expansion at lowest non-trivial order are not displayed; they are evaluated in appendix~\ref{app:tests}.}
\label{fig:z_lambda}
\end{figure}

Let us start by considering the redshift along our rays of light. Figure~\ref{fig:z_lambda} compares the fractional difference between the $z(\lambda)$ relation obtained in our simulation, using the direct methods outlined in section~\ref{subsec:light_propagation_curved_space-time}, with the FLRW relation~$\bar{z}(\lambda)$ obtained by integrating the well-known expression
\begin{equation}
\ddf{\bar{z}}{\lambda} = -(1+\bar{z})^2 H(\bar{z})
\qquad \text{where} \qquad
H(z) = H_0 \sqrt{\Omega\e{m}(1+z)^3+\Omega_\Lambda} \, .
\end{equation}
We display results for both the galaxy simulation (with a compact opaque core), and the halo simulation (with a transparent and diffuse distribution of mass). It can be see that the mean deviation from the predictions of the corresponding FLRW models is less than one part in $10^5$, and decreases as the integration continues. While the numerical values obtained for $\ev{z}_\Omega/\bar{z}-1$ are larger than the error bars displayed, which account for the statistical deviation of numerical results obtained along $10^5$ different geodesics, they are not significantly larger than either the numerical error estimates obtained for our routine (see appendix~\ref{app:tests}) nor the errors that should be expected from neglecting post-post-Newtonian gravitational potentials~\cite{2015PhRvD..91j3532S,2016PhRvD..94b3505S}. Both of these latter sources of error are conservatively estimated to be at the level of about one part in $10^5$.

The results displayed in figure~\ref{fig:z_lambda} lead us to conclude that the deviations of $z(\lambda)$ from their values in a corresponding FLRW universe are no greater than one part in $10^{5}$. This is an intrinsic limit to the accuracy that one could ever hope to obtain from a perturbative approximation of the gravitational field, as used in this study, and in this sense we have saturated the bound on deviations in redshift that can be induced from post-Newtonian gravity. To quantify the effect of inhomogeneity on cosmological redshift one must therefore go to post-post-Newtonian order (i.e. $\mathcal{O}(\epsilon^4)$), and increase the precision of the numerical integrations to a comparable level of accuracy (around one part in $10^{10}$). This is beyond the scope of the present study.

These results should not, of course, be confused with the effects of peculiar motions on the $z=z(\lambda)$ relation. These motions can have a very considerable impact on redshifts~\cite{2012MNRAS.419.1937M,2013PhRvL.110b1302B}, as Doppler effects can contribute significantly to physical observables such as redshift-space distortions. By choosing a set of observers who are as close to comoving as possible, as specified by the four-vector field in equation~(\ref{eq:four_velocity_comoving}), we have explicitly neglected such contributions. In a fully realistic model these effects can, and should, be added to the background cosmological redshift. This can be done in a relatively straightforward way by simply boosting either the observer or the source, but the effects of this are well studied, and would work in exactly the same way here as they do in the standard approach to cosmological modelling. We will therefore not consider them any further.

The negligible effect of inhomogeneity on background redshifts that we have found here can be compared to similar results in the Einstein-Straus Swiss cheese model~\cite{2014JCAP...06..054F}, where deviations from $\bar{z}(\lambda)$ are due to the Rees-Sciama effect~\cite{1968Natur.217..511R}. These models similarly neglect the contribution of peculiar velocities. On the other hand, standard cosmological perturbation theory, LTB, and Szekeres Swiss cheese models display much larger fluctuations in the observed redshifts. In each of these cases the cause of the difference is the peculiar motion about the large-scale average (although the exact models listed here do not require a background in order to be defined). Such effects could be incorporated into the general framework of post-Newtonian cosmological modelling, but would require a more realistic distribution of matter in each cell.

\subsection{Angular distance}
\label{subsubsec:D_A}

\begin{figure}[t!]
\centering
\includegraphics[width=0.49\columnwidth]{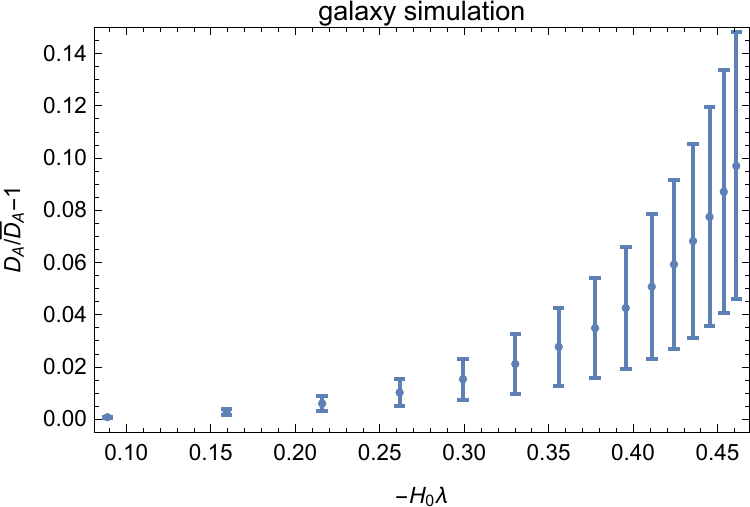}
\hfill
\includegraphics[width=0.49\columnwidth]{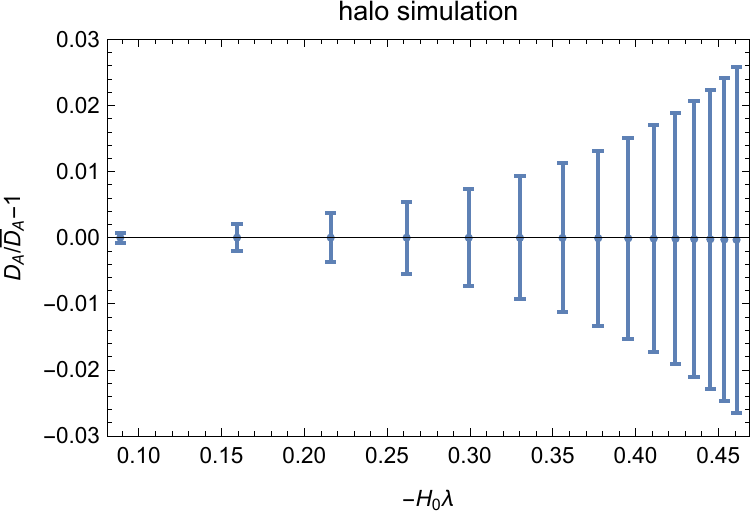}
\caption{The angular diameter distance as a function of affine parameter, $D\e{A}(\lambda)$, as a fraction of the value expected from FLRW cosmology. Dots indicate the statistical mean~$\ev{D\e{A}}_\Omega/\bar{D}\e{A}-1$ from $10^5$ beams of light shot in random directions, and error bars indicate the standard deviation for the same data set. The errors bars are much larger than the estimated numerical error, and correspond to physical phenomena that should have a counterpart in the real Universe.}
\label{fig:DA_lambda}
\end{figure}

We now turn to angular diameter distance measurements. In this case, the FLRW relation between angular distance~$D\e{A}$ and affine parameter~$\lambda$ is the solution of the following differential equation:
\begin{equation}
\ddf[2]{\bar{D}\e{A}}{\lambda} = -4\pi G \rho_0 (1+z)^5 \, \bar{D}\e{A} \, ,
\end{equation}
with $\bar{D}\e{A}(0)=0$ and $\dd\bar{D}\e{A}/\dd\lambda(0)=-1$. In figure~\ref{fig:DA_lambda} we compare the output~$D\e{A}(\lambda)$ we obtain from ray tracing within our simulations to the $\bar{D}\e{A}(\lambda)$ expected from in a comparable FLRW model. Contrary to $z(\lambda)$, the difference can be significant, and in the galaxy simulations is much greater than the estimated numerical error.

In the galaxy simulation the mean distance~$\ev{D\e{A}}_\Omega$, measured across the observer's sky at a given~$\lambda$, is systematically greater than its FLRW counterpart. In other words, light sources are systematically demagnified compared to observations in a homogeneous Universe. This demagnification reaches $\sim 10\%$ for values of $\lambda$ that correspond to $z\sim 1$. The reason for this difference is that light never crosses the opaque matter clumps in these simulations, but instead always propagates through regions of perfect vacuum. This means that the sole source of focussing in the beam is due to the Weyl curvature of the space-time, with no contribution from the Ricci curvature (which vanishes in vacuum). This is the exact opposite of what happens to a beam of light in an FLRW geometry, where Ricci focussing is always non-zero while the Weyl curvature vanishes. The point of relevance here is that the shear that is caused by the Weyl curvature is much less efficient than Ricci curvature at focussing beams of light (except when light passes very close to clumps of matter). It is this lack of focussing that is responsible for the behaviour seen in figure~\ref{fig:DA_lambda}. The dispersion of our data, represented by the error bars in figure~\ref{fig:DA_lambda}, is due to the fact that some light beams pass closer to galaxies than others, and are thus more sheared. Such behaviour is reminiscent of what occurs in Einstein-Straus models~\cite{2014JCAP...06..054F}. These issues will be discussed in more detail below.


In the halo simulation, on the other hand, light sources are not systematically demagnified. In fact, they are even slightly magnified on average (see section~\ref{subsubsec:WKL}). Since the haloes are transparent, the deficit of Ricci focussing that occurs when light propagates through empty space is compensated when the haloes are crossed. As well as this, the fact that the haloes are less compact means that they tend to produce weaker Weyl curvature, meaning that the beams are less sheared. The dispersion of the data, which is slightly smaller than in the galaxy case, is due to the fact that some lines of sight pass through more haloes than others. Unlike redshifts, distance measures are not directly affected by peculiar matter flows, but rather due to gravitational lensing phenomena which is itself related to the local space-time curvature experienced by the beams of light. 

\subsection{Hubble diagram}
\label{sec:hubble}

\begin{figure}[t!]
\centering
\includegraphics[width=0.49\columnwidth]{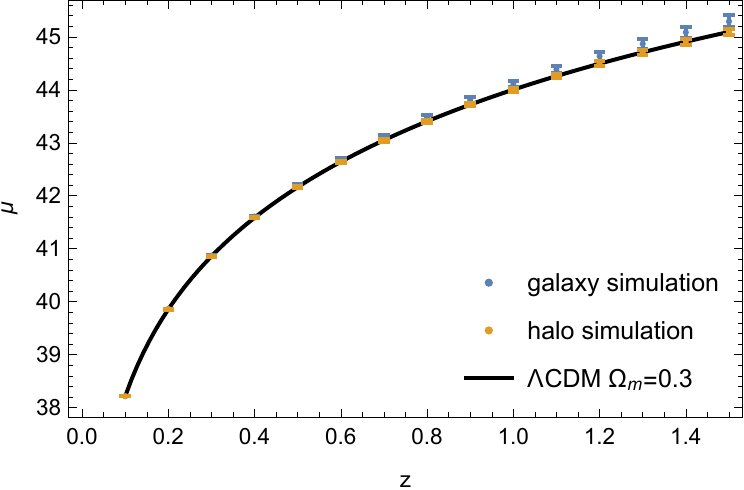}
\hfill
\includegraphics[width=0.49\columnwidth]{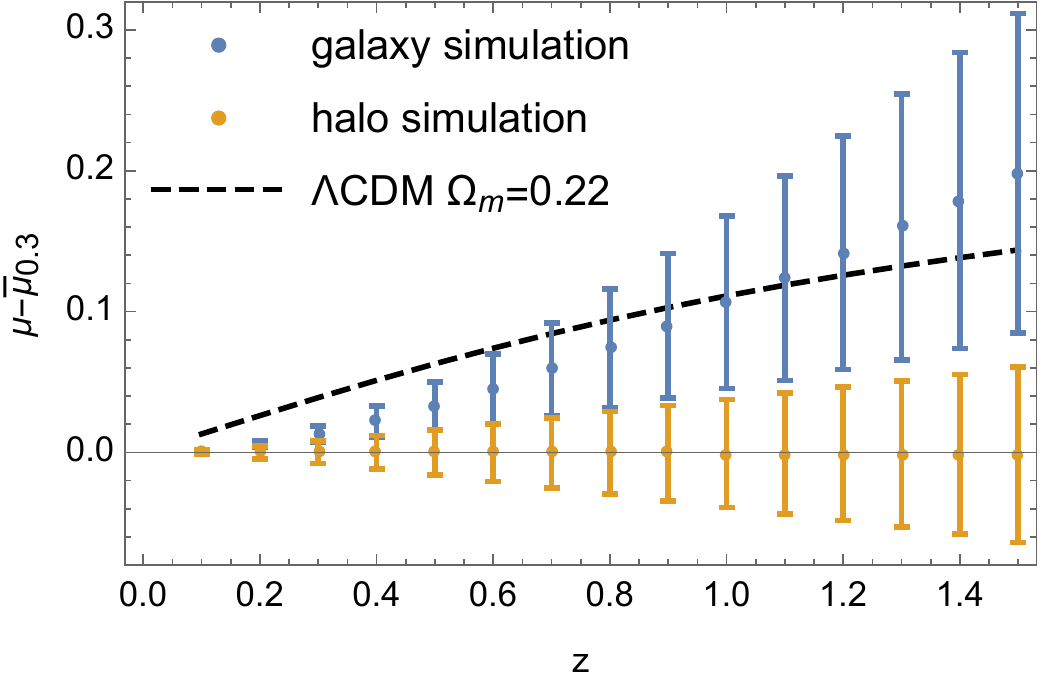}
\caption{\emph{Left panel:} The Hubble diagram constructed in our two post-Newtonian simulations, compared with the fiducial $\Lambda$CDM model constructed with the same cosmological parameters~($\bar{\mu}_{0.3}(z)$ and~$\Omega_{\Lambda}=0.7$). \emph{Right panel:} A difference plot between the simulations and the fiducial $\Lambda$CDM model. The $\Lambda$CDM that best fits the galaxy simulation is also shown. Again, dots indicate the statistical mean of the data set and error bars indicate the dispersion.
}
\label{fig:Hubble_diagrams}
\end{figure}

Let us now combine the results from sections~\ref{subsubsec:D_A} and~\ref{subsubsec:z}, to construct the Hubble diagram that an observer would produce within our model space-time. By convention, it is usual to present this information as a plot of the distance modulus~$\mu$, which is related to the luminosity distance~$D\e{L}$ by
\begin{equation}
\mu \define 5\log \pa{ \frac{D\e{L}}{10\U{pc}} } \, .
\end{equation}
In figure~\ref{fig:Hubble_diagrams} we compare the Hubble diagrams generated from our simulations, within a post-Newtonian model, to those expected in a comparable FLRW model. As expected from section~\ref{subsubsec:D_A}, the differences are most pronounced in the case of galaxy simulations. This follows immediately from the result that redshifts are largely unaffected by the inhomogeneity, leading to
%
%
%
$
\delta D\e{A}(z) 
\define D\e{A}[\lambda(z)] - \bar{D}\e{A}[\bar{\lambda}(z)] 
\approx \delta D\e{A}[\bar{\lambda}(z)] \, .
$
Note, however, that this approximate equality is no longer true when second-order terms are taken into account. Such terms can bias the Hubble diagram in a non-trivial way, with the quantitive magnitude of the effect depending on the particular distance measure being considered (luminosity distance, luminous intensity, distance modulus, etc.)~\cite{2016MNRAS.455.4518K,2015JCAP...07..040B,2016arXiv161203726F}.

\begin{figure}[b!]
\centering
\includegraphics[width=0.5\columnwidth]{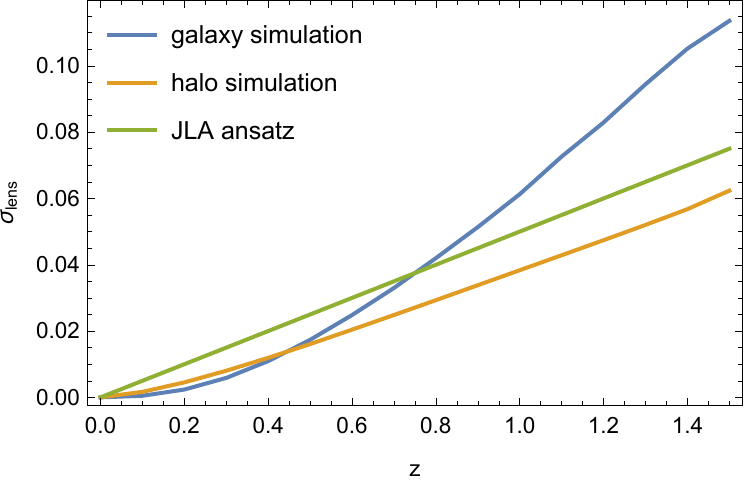}
\caption{Dispersion of the Hubble diagram~$\sigma\e{lens}$ due to gravitational lensing in each of our two simulations, compared with the ansatz of the joint light-curve analysis~\cite{2014A&A...568A..22B}.}
\label{fig:dispersion_Hubble_diagram}
\end{figure}

It is interesting to note that in the galaxy simulation the effect of inhomogeneity on the Hubble diagram acts to bias the inferred amount of dark energy. That is, if we fit the data from our galaxy simulation to a spatially flat  $\Lambda$CDM model by minimizing the $\chi^2$ statistic,
\begin{equation}
\chi^2(\Omega\e{m})
\define
\sum_{i=1}^{15} \pac{ \frac{\ev{\mu\e{sim}}_\Omega(z_i)-\mu_{\rm\Lambda CDM}(z_i;\Omega\e{m})}{\sigma(z_i)} }^2 \, ,
\end{equation}
then we find that best-fitting model has $\Omega\e{m}=0.22$, rather than the value of $0.3$ that would be obtained if one had direct access to the background rate of expansion. In performing this fit we have taken $z_i=0.1 i$ and $\sigma^2(z_i)=\sigma\e{int}^2+\sigma\e{lens}^2(z_i)$, where $\sigma\e{int}=0.1$ is the intrinsic dispersion that is expected to encapsulate the astrophysics of supernovae in real data~\cite{2011ApJS..192....1C} and $\sigma\e{lens}$ is the dispersion due to gravitational lensing (corresponding here to the error bars in figure~\ref{fig:Hubble_diagrams}). This is comparable to similar results found in the Einstein-Straus Swiss cheese model~\cite{2013PhRvD..87l3526F}. It implies that if one fits the observed Hubble diagram in such a universe, by wrongly assuming that it is homogeneous and isotropic, then one overestimates its actual dark energy content by about $\Delta \Omega_{\Lambda}=0.08$. When applied to supernova data, this effect tends to improve agreement with \textsl{Planck}~\cite{2013PhRvL.111i1302F} on $\Omega\e{m}$.

Finally, let us close this section with a remark on the scatter of supernova data due to gravitational lensing in the joint light-curve analysis (JLA)~\cite{2014A&A...568A..22B}. The results of the JLA, which assume~$\sigma\e{lens}(z)=0.055 z$, are shown in figure~\ref{fig:dispersion_Hubble_diagram} together with our numerical results from ray tracing. The JLA ansatz can be seen to lie somewhere between our halo model and galaxy model. The shape of the JLA ansatz is closer to the halo case, which starts to become linear around $z=0.4$, while in the galaxy case the behaviour is more complicated. This is not surprising as the JLA ansatz is motivated by standard cosmological perturbation theory, where lensing is mostly due to fluctuations in the Ricci curvature (as is the case in our halo simulation). This can be seen as a confirmation of the JLA approach, but also warns that more complicated effects can occur in a universe that contains high-density compact astrophysical structures. However, the reader may note that the precise values of $\sigma\e{lens}$ in our simulations must be taken with some caution as they strongly depend on the parameters of the model, as discussed in section~\ref{subsec:stochastic_lensing}.

\section{Discussion}
\label{sec:discussion}

In section~\ref{sec:results} we presented the basic results from our ray-tracing code, in terms of redshifts and distance measures. The difference between the averages of some of these data sets, and the expectations from FLRW models, were sometimes seen to be significant. In addition, the variance within each data set has the potential to allow interesting information about the structure in the Universe to be extracted from (for example) supernova observations. It is therefore of considerable interest to be able to relate these results to the underlying features of the model, in order to be able to understand how different features of the each configuration affects cosmological observations, and to be able to extrapolate these results from our idealized models to the real Universe.

To this end, in this section, we will interpret the results from section~\ref{sec:results} in terms of some of the most prominent frameworks that have been constructed to understand the effects of inhomogeneity on the propagation of light. In particular, we will investigate (i) the theorem developed by Weinberg and Kibble \& Lieu that relates the mean of observables to FLRW expectations, (ii) the stochastic lensing formalism that aims to model the statistical distribution of observations made along many lines of sight, and (iii) the relationship with the empty beam approximation along special lines of sight, as recently studied using full numerical relativity. This will allow us to probe the regimes in which these constructions and approximations are valid, as well as provide ways to understand the results of the previous section within well-established frameworks.

\subsection{Weinberg-Kibble-Lieu theorem}
\label{subsubsec:WKL}

In 1976 it was argued by Weinberg that gravitational lensing due to inhomogeneity in the Universe does not change the average relation between luminosity distance and redshift~\cite{1976ApJ...208L...1W}. However, it was later realised that this very general statement could not be true; Weinberg overlooked the importance of choosing the observable that one averages, as well as the importance of the choice of averaging procedure itself. That is, averaging luminosity distances is not equivalent to averaging magnitudes or luminous intensities, and averaging over angles is not equivalent to averaging over sources or area~\cite{2013PhRvL.110b1301B,2013JCAP...06..002B,2015MNRAS.454..280K,2016MNRAS.455.4518K,2015JCAP...07..040B,2016arXiv161203726F}. A more detailed and accurate statement was proposed in 2005 by Kibble \& Lieu~\cite{2005ApJ...632..718K}, which stated that it is the directional average of the angular diameter distance squared on surfaces of constant $\lambda$ that is unaffected by the presence of matter inhomogeneities. Mathematically, this can be expressed as follows:
\begin{equation}\label{eq:WKL}
\ev{D\e{A}^2(\lambda)}_\Omega \define \frac{1}{4\pi} \int_{\lambda=\cst} D\e{A}^2 \, \dd\Omega = \bar{D}\e{A}^2(\lambda) \, .
\end{equation}
Equivalently, one can say that the total area of surfaces of constant $\lambda$,~$A(\lambda)=4\pi\ev{D\e{A}^2(\lambda)}_\Omega$, is unchanged by gravitational lensing. We shall call this property the Weinberg-Kibble-Lieu (WKL) theorem. If one further assumes that the $z(\lambda)$ relation is not significantly affected by inhomogeneities, the WKL theorem can be re-written in terms of surfaces of constant $z$.

\begin{figure}[t!]
\centering
\includegraphics[width=0.47\columnwidth]{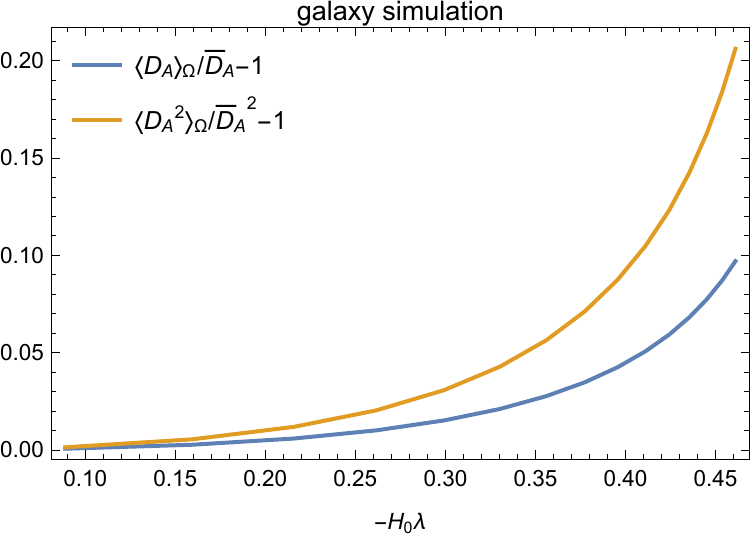}
\hfill
\includegraphics[width=0.5\columnwidth]{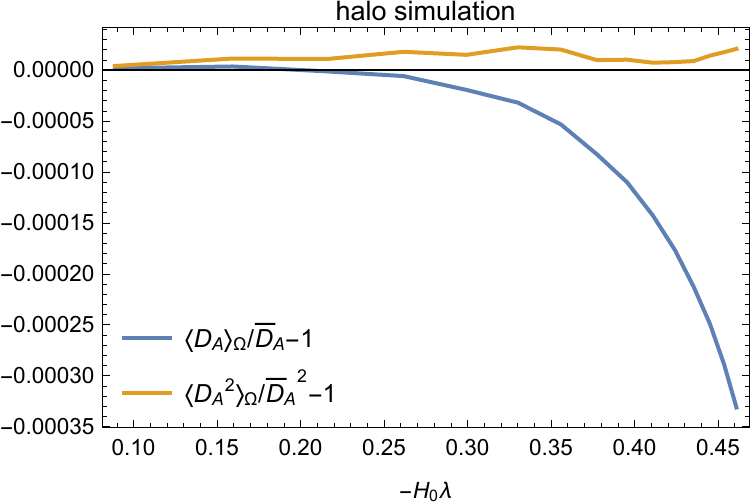}
\caption{Evaluation of the Weinberg-Kibble-Lieu theorem, according to which~$\ev[1]{D\e{A}^2(\lambda)}_\Omega=\bar{D}\e{A}^2(\lambda)$, with ray-tracing in post-Newtonian simulations.}
\vspace{-0.2cm}
\label{fig:DA_DA2}
\end{figure}

We can now test the validity of eq.~\eqref{eq:WKL} in our PN simulations, by comparing $\ev{D\e{A}}_\Omega$ and $\ev{D\e{A}^2}_\Omega$ to their FLRW counterparts. The results are shown in figure~\ref{fig:DA_DA2}, for both the galaxy and halo simulations. The WKL theorem is clearly violated in the case of the galaxy simulation, while it appears an extremely good approximation in the case of the halo simulation. This difference is essentially due to the opacity of the galaxies, which means that the light rays cannot sample every point in space. The region that is excised in the galaxy simulation is the region of highest Ricci curvature, meaning that the average density of matter along the allowed lines of sight is lower than the cosmological average (or strictly zero, in the idealized simulation we have performed). This biases the average of the inferred measures of distance, and causes the discrepancy that can be seen in figure~\ref{fig:DA_DA2}. The  halo simulation, on the other hand, has no regions of space that the rays of light are forbidden from entering. The average density they experience is therefore much closer to the cosmological average. This is the essential reason why the WKL theorem works so well in the halo case\footnote{The reader may note that the accuracy of these results is much greater than the random error on any individual trajectory; we attribute this to the error on the mean decreasing as the inverse square root of the number of trajectories in the sample (which is $10^5$ here).}.

The logic above is not only applicable to our post-Newtonian models alone, but should apply to every model in which opaque structures are present (including the real Universe). In particular, similar behaviour has been observed in the context of Einstein-Straus Swiss cheese models~\cite{2014JCAP...06..054F}. A further property that causes violation of the WKL theorem, that is not present in our models but may exist in more general setups, is statistical anisotropy~\cite{2016JCAP...06..008F}. This can cause additional Weyl lensing, which means that $\ev{D\e{A}^2(\lambda)}_\Omega \neq \bar{D}\e{A}^2(\lambda)$. Finally, the averaging over angles~$\ev{\ldots}_\Omega$ that appears in eq.~\eqref{eq:WKL} is not always the relevant averaging procedure to perform for every cosmological observable; it is the natural notion of averaging that one would use for full sky observations such as the CMB~\cite{2015JCAP...06..050B, 2015arXiv150706590L}, but is not the type of averaging that one usually uses when constructing a Hubble diagram from observations~\cite{2016arXiv161203726F}. For this latter case it would be more natural to perform an average over sources, which could lead to violations of the relations one might expect from naive extrapolation.

\subsection{Comparison with stochastic lensing}
\label{subsec:stochastic_lensing}

A useful formalism for modelling the effects of small-scale inhomogeneities on the propagation of narrow light beams has recently been constructed by modelling lensing events as stochastic processes~\cite{2015JCAP...11..022F}. In a nutshell, this approach consists of modelling the fluctuations in the Ricci and Weyl focussing scalars as white noise. The moments of the distribution of angular diameter distance measures can then be calculated, and compared to observations and ray-tracing models. It is instructive to compare this approach to the results we obtained in section~\ref{sec:results}, in order to evaluate its efficacy, and in order to determine the features of the real Universe that it should be expected to faithfully capture.

The first step in the stochastic formalism of ref.~\cite{2015JCAP...11..022F} is to decompose Ricci focussing term from equation~(\ref{tidal}) as~$\Ricfoc=\ev{\Ricfoc}+\delta\Ricfoc$, where $\ev{\Ricfoc}=-4\pi G (1+z)^2 \ev{\rho}$ is the mean Ricci focusing associated with the mean matter density~$\ev{\rho}$ encountered by the light beam, while~$\delta\Ricfoc$ encodes fluctuations about it. There is no need for such a decomposition of~$(\Weylfoc_1,\Weylfoc_2)$, as in a statistically isotropic universe $\ev{\Weylfoc_A}=0$. The stochastic processes, $\delta\Ricfoc$ and $\Weylfoc_A$, are then characterised by covariance functions~$C_\Ricfoc$ and~$C_\Weylfoc$, defined by
\begin{align}
\ev{\delta\Ricfoc(\lambda) \delta\Ricfoc(\lambda') } &= C_\Ricfoc(\lambda) \delta(\lambda-\lambda') \, , \\[10pt]
\ev{\Weylfoc_A(\lambda) \Weylfoc_B(\lambda') } &= C_\Weylfoc(\lambda) \delta_{AB} \delta(\lambda-\lambda') \, ,
\end{align}
where $\delta(\lambda-\lambda')$ is the Dirac delta function and $\delta_{AB}$ is the Kronecker delta. From these assumptions, it is now possible to calculate the mean and variance of the angular diameter distance along different lines of sight.

In particular, the lowest-order contribution to the average of the angular distance~$\ev{D\e{A}}$ due to shear can be written as~\cite{2015JCAP...11..022F}
\begin{equation}\label{eq:pKDR}
\ev{D\e{A}(\lambda)}
= D_0(\lambda) \left( 1+  2 \int_0^\lambda \frac{\dd \lambda_1}{D_0^2(\lambda_1)}
	\int_0^{\lambda_1} \frac{\dd \lambda_2}{D_0^2(\lambda_2)}
	\int_0^{\lambda_2} \dd \lambda_3 \,
		D_0^4(\lambda_3) C_\Weylfoc(\lambda_3)
+ \mathcal{O}(C_\Weylfoc^2) \right) \, ,
\end{equation}
where $D_0$ denotes the angular diameter distance in a homogeneous space with $\ev{\Ricfoc}$ and $\Weylfoc=0$.
Next, the variance of angular diameter distances, $\varDA(z) \define \ev{D\e{A}^2(z)}_\Omega - \ev{D\e{A}(z)}^2_\Omega$, can be shown to satisfy the following differential equation~\cite{2015JCAP...11..022F}:
\vspace{0.1cm}
\begin{multline}\label{eq:standard_deviation_DA}
(\varDA)'''
+ \pa{\frac{H'}{H} +\frac{6}{1+z}} (\varDA)''
+ \Bigg[ \frac{H''}{H}+\pa{\frac{H'}{H}}^2 
			+ \frac{8}{1+z} \frac{H'}{H}
			+ \frac{6}{(1+z)^2}
			- \frac{4\ev{\Ricfoc}}{(1+z)^4 H^2} 
		\Bigg] (\varDA)' \\
+ \frac{2\ev{\Ricfoc}'}{(1+z)^4 H^2} + \frac{ 4C_\Weylfoc - 2C_\Ricfoc }{(1+z)^6 H^3} \, \varDA \\[10pt]
=
\frac{2 D_0^2 C_\Ricfoc}{(1+z)^6 H^3}
+
\frac{6}{(1+z)^6 H^3 D_0^4} \int_0^z \frac{\dd z_1}{(1+z_1)^2 H D_0^2}
\pac{ \int_0^{z_1} \dd z_2 \; \frac{2D_0^4 C_\Weylfoc}{(1+z_2)^2 H}}^2 \, , \\ 
\end{multline}
where it has been assumed that $z(\lambda)$ is the same as in the corresponding FLRW model, and where a prime denotes a derivative with respect to $z$. It can be seen that both the Ricci term and an integrated Weyl term act as sources for $\varDA$.

It now remains to determine the covariance functions~$C_\Weylfoc$ and~$C_\Ricfoc$. In a universe randomly filled with spherical opaque clumps of matter with mass~$M$ and size~$R$, the associated Weyl covariance is found to be~\cite{2015JCAP...11..022F}
\begin{equation}
C_\Weylfoc \approx \frac{3g}{2} \, H_0^2 \Omega\e{m} (1+z)^6 \, ,
\end{equation}
where $g=GM/R^2$ is the gravitational acceleration at the surface of the clump. This result was originally derived in the context of Swiss cheese models, but can be easily generalised to the present case. Finally, it can be shown that for the galaxy simulations we have~$C_\Ricfoc=0$, while for halo simulations we have
\begin{equation}
C_\Ricfoc 
\approx \frac{72}{5} \frac{L_0^3}{R^2} H_0^4 \Omega\e{m} (1+z)^6 \, .
\end{equation}
We now have all the information required to compute~$\ev{D\e{A}(\lambda)}$ and~$\varDA$ for our two simulations, and to compare these to our results from ray tracing.

\begin{figure}[t!]
\centering
\includegraphics[width=0.49\columnwidth]{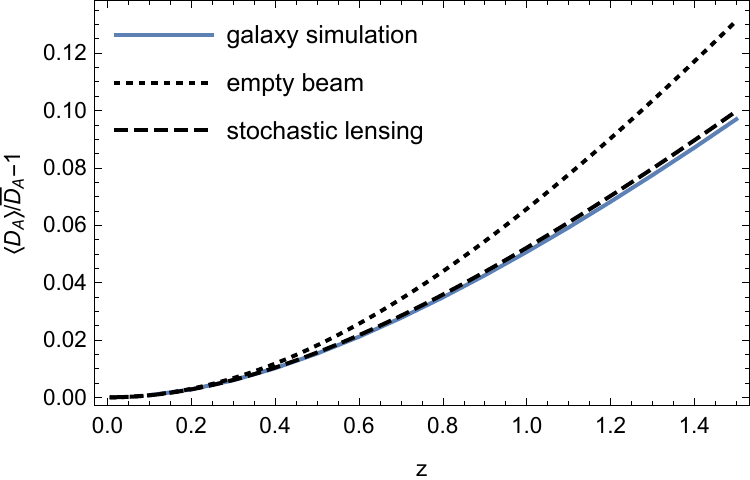}
\hfill
\includegraphics[width=0.49\columnwidth]{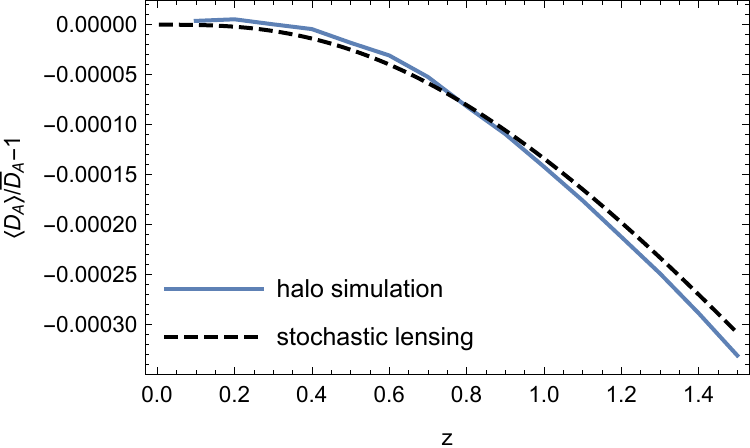}
\caption{Comparison between the sky-averaged angular diameter distance~$\ev{D\e{A}}_\Omega$ in our post-Newtonian simulations, and the predictions of the stochastic lensing formalism. In the galaxy case (left panel), we have also plotted the results one would obtain from using the empty beam approximation.}
\label{fig:bias_DA}
\end{figure}

In figure~\ref{fig:bias_DA} we compare the mean angular diameter distance from our ray tracing experiment to the results from the stochastic lensing formalism. We see that the prediction from equation~\eqref{eq:pKDR} reproduce the results of our simulations very accurately. In the galaxy case we have also indicated the results that one would obtain from the empty-beam approximation, in which~$\Ricfoc=0$. We see that this empty beam approximation, although not very good, is still better than FLRW at modelling the results from ray tracing. This confirms that, in the case of our galaxy simulations, most of the departure from FLRW is caused by the deficit of Ricci lensing due to the opacity of the matter clumps. When the stochastic shear corrections are added to $D_0$, the numerical results are recovered to high accuracy. In the halo case, where the FLRW model is already a good model for $\ev{D\e{A}}_\Omega$, the stochastic shear correction from eq.~\eqref{eq:pKDR} can be seen to improve the fit even further.

\begin{figure}[t!]
\centering
\includegraphics[width=0.49\columnwidth]{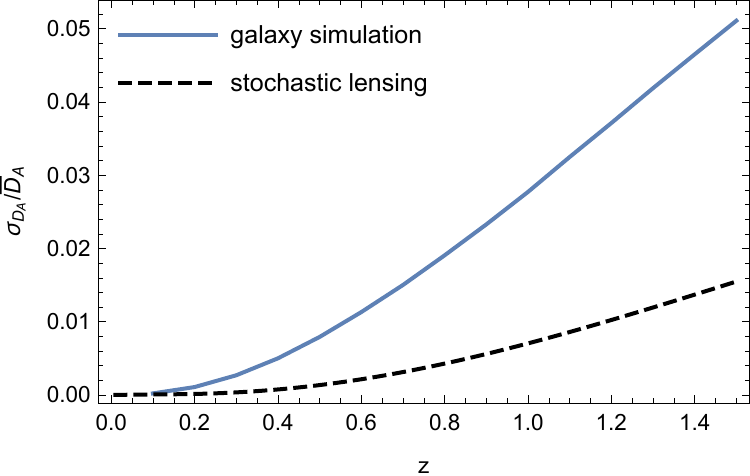}
\hfill
\includegraphics[width=0.49\columnwidth]{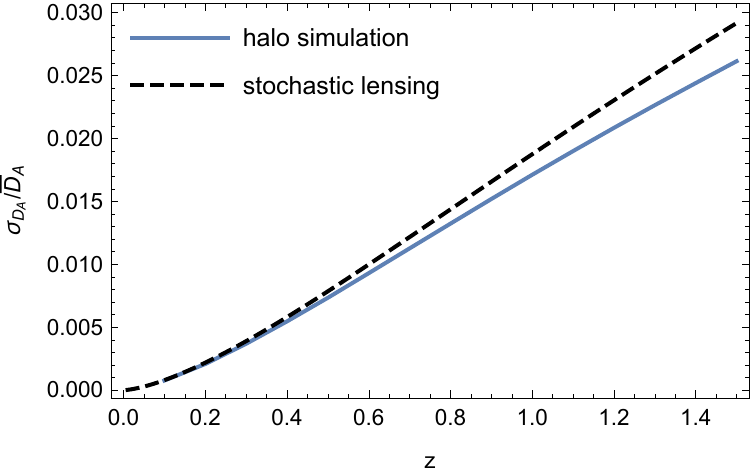}
\caption{Comparison between the dispersion~$\sigma_{D\e{A}}$ of the angular distance in our post-Newtonian simulations, and the prediction from the stochastic lensing formalism.}
\label{fig:dispersion_DA}
\end{figure}

Finally, in figure~\ref{fig:dispersion_DA} we compare the variance of our ray tracing results with the results obtained from stochastic lensing. We see that equation~\eqref{eq:standard_deviation_DA} provides an excellent description of the halo simulations, but fails in the galaxy case. The inability of stochastic lensing to predict the correct variance of the angular distance in a universe where light passes through vacuum regions is caused by the strong non-Gaussianity of $\Weylfoc$, which cannot be reasonably modelled as a white noise. This supports similar findings in the case of the Einstein-Straus Swiss cheese model~\cite{2015JCAP...11..022F}.

\subsection{Comparison with numerical relativity}

\begin{figure}[b!]
\centering
\includegraphics[width=0.45\columnwidth]{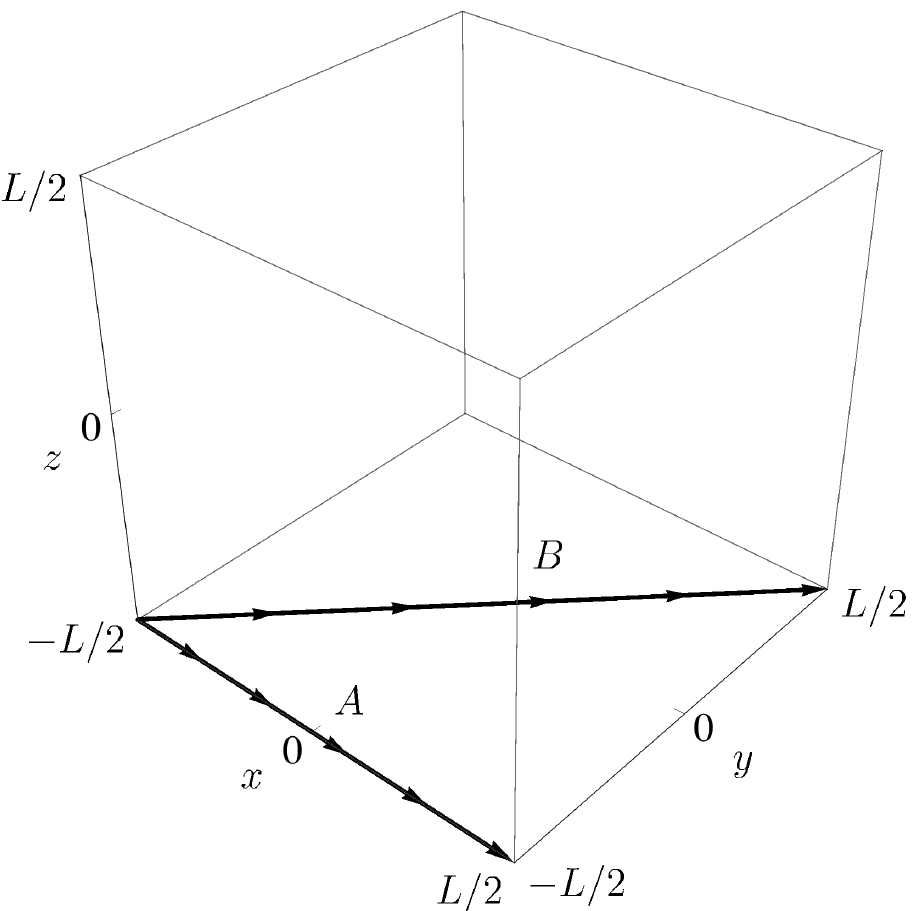}
\caption{Light rays $A$ and $B$ within a cubic lattice cell. This figure has been adapted from ref.~\cite{Bentivegna:2016fls}.}
\label{fig:geodesics_A_B}
\end{figure}

A recent field of study, within the field of inhomogeneous cosmology, is the study of models created using the tools from numerical relativity~\cite{Bentivegna:2012ei, Bentivegna:2013xna, Bentivegna:2013jta, Korzynski:2015isa, Yoo:2012jz, Yoo:2013yea, Yoo:2014boa, 2015PhRvL.114e1302A, 2016NatPh..12..346A, 2016JCAP...07..053A, 2016PhRvD..93l4059M, 2016ApJ...833..247G, 2016PhRvL.116y1302B, 2016arXiv161103437D}. The prospect of creating high-accuracy inhomogeneous cosmological space-times without any continuous symmetries is a highly exciting one, and presents the prospect of testing a number of the ideas around the possible consequences of inhomogeneity in cosmology in a rigorous and well-defined way. Nevertheless, the construction of numerical cosmological models is not without its difficulties. Numerical artefacts in these solutions do not dissipate in the same way that they do in asymptotically flat space-times, and the amount of computational time required to calculate the space-time geometry (and observables within it) is not trivial. 

These facts make comparison between numerical relativity and approximation schemes potentially useful for those who are trying to eliminate the effects of numerical errors and artefacts in their simulations. In the future, such a comparison may also provide a way for us to try and evaluate the extent to which approximation schemes such as ours can be used as valid approaches for trying to create more versatile frameworks for capturing the lead-order effects of inhomogeneity. To these ends, we will compare our ray-tracing results to the only results of calculating observables within a numerical cosmological model that are currently available; those of ref.~\cite{Bentivegna:2016fls}. Although the two approaches to cosmological modelling are formally quite different, the properties of the physical configuration they are trying to simulate are very similar: In both cases indeed the universe is periodic and made of compact matter clumps.

As in our models, the authors of ref.~\cite{Bentivegna:2016fls} considered a universe tiled with cubic lattice cells. They then consider two fiducial light beams, $A$ and $B$, as depicted in figure~\ref{fig:geodesics_A_B}. These beams are very special, in that they are stable during their propagation; the periodicity of the model implies that neither of them can be deflected by the masses at the centre of the cells. This is guaranteed by the fact that reflection symmetric surfaces are always totally geodesic, and that each beam lies at the intersection of at least two reflection symmetric surfaces. Along each of their trajectories they then computed the $D\e{L}(z)$ relation, which is a direct observable for anyone inside the lattice.

In ref.~\cite{Bentivegna:2016fls}, the authors then compared the results of numerically integrating the optical equations in their numerical space-time with the following well-known cosmological models:
\begin{description}
\item[(i) Einstein-de Sitter (EdS):] An FLRW model with~$\Omega\e{m}=1$ and $\Omega_K=\Omega_\Lambda=0$;
\item[(ii) $\Lambda$CDM:] The concordance FLRW model with~$\Omega\e{m}=0.3$, $\Omega_K=0$, and $\Omega_\Lambda=0.7$;
\item[(iii) The Milne universe:] An FLRW universe with $\Omega\e{m}=0$, $\Omega_K=1$, and $\Omega_\Lambda=0$;
\item[(iv) The empty-beam approximation (EBA):] A cosmology with~$D\e{L}=-(1+z)^2\lambda$, and the EdS relation between redshift and affine parameter~$z(\lambda)$.
\end{description}
They found that the last of these approaches is the best (and a good) fit to their numerical results. Initially, this may seem at odds with the results we presented in the left panel of figure~\ref{fig:bias_DA}, but we will show it is entirely consistent.

\begin{figure}[t!]
\centering
\begin{minipage}{0.55\columnwidth}
\includegraphics[width=\columnwidth]{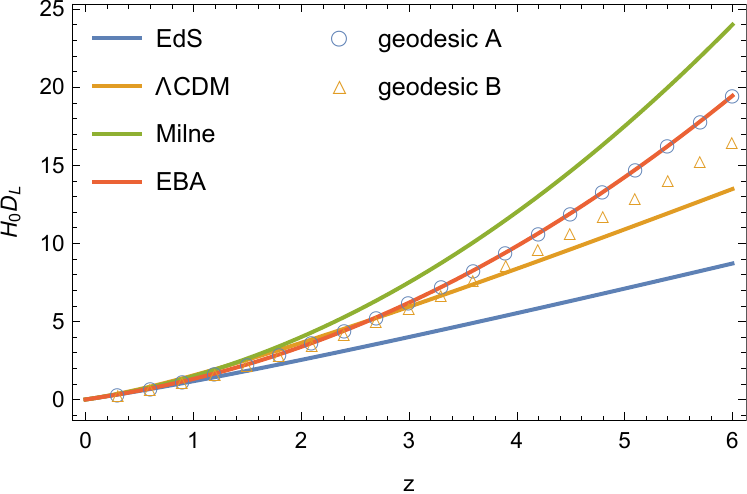}
\end{minipage}
\hfill
\begin{minipage}{0.42\columnwidth}
\includegraphics[width=\columnwidth]{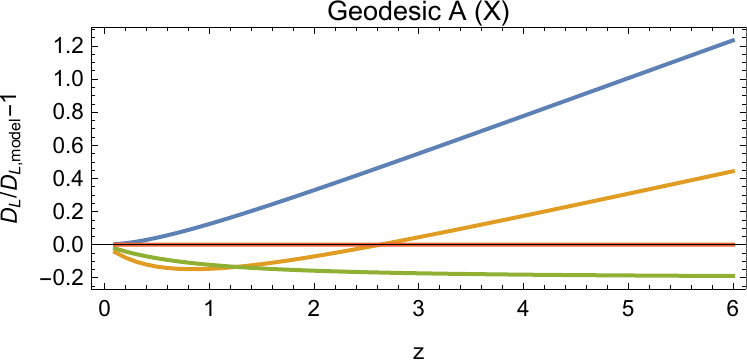}\\
\includegraphics[width=\columnwidth]{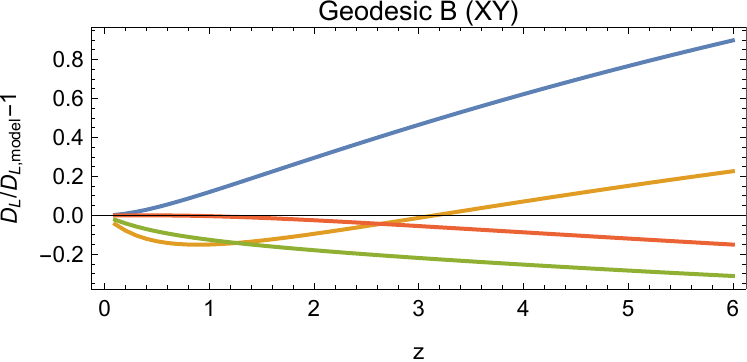}
\end{minipage}
\caption{Comparison between the luminosity distance-redshift relation~$D\e{L}(z)$ along the light rays $A$ and $B$, as depicted in figure~\ref{fig:geodesics_A_B}, and the following four models: (i) Einstein-de Sitter (EdS), (ii) $\Lambda$CDM, (iii) Milne, and (iv) the empty-beam approximation (EBA) in an EdS background.}
\label{fig:comparison_Bentivegna}
\end{figure}

In figure~\ref{fig:comparison_Bentivegna}, we present the results of considering the special light beams, $A$ and $B$, in our post-Newtonian simulations (without the cosmological constant, in order to facilitate a direct comparison with ref.~\cite{Bentivegna:2016fls}). Indeed, it can be readily observed that the EBA does indeed provide the best-fitting $D\e{A}(z)$ along each of these beams in our simulations, just as it does in the numerical relativity simulations of ref.~\cite{Bentivegna:2016fls}. In fact, in the case of beam $A$ this result is exact, as along the edge of a cell tidal forces must be exactly zero due to discrete rotational symmetry. Furthermore, the beam does not encounter any form of matter along the cell edge, so both Ricci and Weyl lensing terms must vanish (as assumed in the EBA).  For beam $A$, the apparent discrepancy with the results presented in figure~\ref{fig:bias_DA} can now be seen to be entirely due to the very special choice of geodesics $A$, which is not representative of the experience of a typical ray of light within the model. In particular, beam $A$ experiences a drastically lower rate of shear than almost all other trajectories.

Let us now consider beam $B$, which is slightly focused with respect to the EBA. This is because along $B$ the tidal forces due to the masses above and below are not cancelled by the masses on the sides, which are further away. This beam is thus consistently sheared along the $z$ axis, which causes a correction with respect to the EBA. It is interesting to note that, even though shear is consistently produced along the same axis, the difference between $D\e{A}(z)$ along the trajectory B and the EBA remains very small ($5\times 10^{-5}$ at $z\sim1$). This can be compared with the left panel of figure~\ref{fig:bias_DA}, where the difference between $\ev{D\e{A}}_\Omega$ and the EBA is at the level of one percent at the same redshift. We attribute this to the fact that the beam is always far from the central mass of each cell it passes. Consequently, this means that the majority of the shear corrections in $\ev{D\e{A}}_\Omega$ must come from very close encounters with the central masses, and not from special trajectories where the contributions to shear are all cumulative. We conclude from this that the results presented in section~\ref{sec:results} are not artefacts of the lattice structure of the model, but should be quite robust under changes of configuration.

\section{Conclusions}
\label{sec:conclusion}

In this paper we have ray traced through a set of models that were created to simultaneously model both large-scale expansion and non-linear structure. We have found that the redshifts that an observer in such a space-time would model are very close to those of an FLRW model with the same amount of total mass and dark energy. This result was found to be independent of the way in which matter is distributed within each of the primitive cells of our lattice model, and whether or not the matter is opaque or transparent. On the other hand, we have found that angular diameter and luminosity distances can differ considerably from the predictions of FLRW cosmology, as long as the mass is clumped into compact and opaque non-linear structures. In the extreme case, when all matter is clustered into high-density galaxy-bulge-sized objects we find that the difference can be as much as 10\%, with objects appearing dimmer at the same redshift in the inhomogeneous cosmology. This difference drops dramatically when the matter is taken to be dispersed and transparent, however, and becomes less that 1\% when dark matter halo-sized objects are being modelled.

When the results from considering dense opaque objects are fitted to FLRW cosmological models, we find that the best fitting spatially flat model has $\Omega_{\Lambda}=0.78$ when the actual value in the model is $\Omega_{\Lambda}=0.7$. This constitutes a bias of more than $10\%$ in the estimation of this crucial cosmological parameter, compared to the perfectly homogeneous and isotropic case, which is a considerable effect. Beyond the difference in the mean, the inhomogeneities also add a variance to the distribution around the mean. In the case of clumped opaque matter this is at the level of about $\pm 5\%$ for cosmologically interesting distances, while for diffuse transparent structures it is more like $\pm 3\%$. These are considerable deviations, that could in principle be used to extract information about inhomogeneity from the distribution of, for example, supernova observations.

By comparing our results to the expectations from the Weinberg-Kibble-Lieu theorem, we find that when the matter is transparent and diffuse the relevant averages in the inhomogeneous universe correspond to the FLRW values in a very precise way. When matter is opaque and compact, on the other hand, there is no such correspondence, and the averages of the relevant optical probes in the inhomogeneous universe do not correspond to FLRW values in any obvious way. We find that the recently developed stochastic approach to lensing is excellent at modelling the average of the luminosity distances that we calculate in both the opaque and transparent distributions of matter. This formalism is, however, rather less good at predicting the variance around the mean in the presence of opaque clusters of matter. Finally, we calculate observables along the special curves that have recently been considered in the context of numerical relativity simulations of inhomogeneous cosmological models. Our results are free from the oscillatory numerical artefacts that exist in these models, and could serve as a guide for the expected results that could be obtained once they have been brought under control.

The models that we have been considering should not, of course, be considered to be fully realistic representations of the actual Universe. This is primarily due to the periodicity that is assumed on large scales, in order to make their construction a tractable problem. While the real Universe does appear to display a homogeneity scale, and while similar periodicity assumptions are made in standard $N$-body simulations, this does not imply the stronger assumption of reflection symmetry that we have assumed. This means that the post-Newtonian cosmologies we have considered should be thought of as toy models. Nevertheless, these model do retain the distinct advantage of allowing relativistic effects to be consistently included in cosmology in the presence of non-linear structures, and in this regard they are much more flexible than other comparable approaches (such as Swiss cheese models or black hole lattices).

Due to the idealised nature of our model, we do not expect the $10\%$ bias in $\Omega_{\Lambda}$ discovered in our galaxy simulations to be fully representative of the value of the corresponding effect in the real Universe. This is because the real Universe contains both opaque galaxy bulges and dark matter haloes (as well as various other astrophysical matter). In this sense, the galaxy and halo simulations we have performed should be considered as extreme examples that provide upper and lower bounds on the magnitude of the effect that will occur in the real Universe. Given that we have no reason to expect the periodicity of our model to introduce any further biases beyond those discussed above, and given there is more dark matter than visible matter in the Universe, we estimate that in reality the astronomical selection effects should bias the estimation of $\Omega_{\Lambda}$ at about the level of $\sim 1\%$. This is a significant effect, in the age of precision cosmology.

In subsequent work we will consider the detailed statistical behaviour of the observables of gravitational lensing, such as optical shear and distance measures, as well as including more realistic distributions of matter. A first step in doing this would be to include opaque galaxy cores and, e.g., a Navarro-Frenk-White profile. A second step would be to arrange these haloes within a cell according to the actual matter power spectrum, and allowing for their peculiar motion under gravity. This should provide further insight into the biases that inhomogeneous structures can cause, as well as allowing interesting questions about the formation of caustics and the statistics of optical shear and distance measures to be addressed. If caustics form in any considerable number, then they could have profound consequences for both the construction of Hubble diagrams and CMB observations~\cite{Ellis:1998ha}.


\section*{Acknowledgements}

PF thanks Queen Mary University of London for its great hospitality at several stages of this project. VAAS would like to thank John Ronayne, Pedro Carrilho, Mike Cole, Cozmin Timis and Alex Owen for very helpful discussions and advice. We also thank Eloisa Bentivegna, Miko\l{}aj Korzy\'{n}ski, Syksy R\"{a}s\"{a}nen, Ian Hinder and Jean-Philippe Uzan for comments on the draft version of this article. VAAS and TC are both supported by the STFC. PF is supported by the Swiss National Science Foundation.


\appendix

\section{Perturbed geometric quantities}
\label{app:geometry}

At order $\eps^2$ in the post-Newtonian expansion, the metric of space-time is given in equation~(\ref{eq:metric}). The perturbed Christoffel symbols associated with this geometry are then
\begin{align}
\Gamma\indices{^\mu_t_t}, \Gamma\indices{^t_\mu_t} &=-\Phi_{,\mu} + \mathcal{O}(\eps^3) \, , \\
\Gamma\indices{^\rho_\mu_\nu} &= \delta_{\rho\mu}\Psi_{,\nu} + \delta_{\rho\nu} \Psi_{,\mu} - \delta_{\mu\nu}\Psi_{,\rho} + \mathcal{O}(\eps^3) \, ,\\
\Gamma\indices{^t_t_t}, \Gamma\indices{^\mu_\nu_t}, \Gamma\indices{^t_\mu_\nu} &= \mathcal{O}(\eps^3) \, , \\
\Gamma\indices{^\mu_t_t} &= \mathcal{O}(\eps^4) \, ,
\end{align}
and the Riemann tensor reads
\begin{align}
R_{t \mu t \nu} &= -\Phi_{,\mu\nu} + \mathcal{O}(\eps^4) \, ,\\
R_{t \mu \nu \rho} &= \mathcal{O}(\eps^3) \, ,\\
R_{\mu\nu\rho\sigma} &= \delta_{\mu\sigma} \Psi_{,\nu\rho} 
+ \delta_{,\nu\rho}  \Psi_{\mu\sigma}
- \delta_{\mu\rho} \Psi_{,\nu\sigma} 
- \delta_{,\nu\rho} \Psi_{\mu\sigma}
+\mathcal{O}(\eps^4) \, .
\end{align}
These are all the components required for our numerical integrations.

\section{Tests in de Sitter space-time}
\label{app:tests}

\begin{figure}[t!]
\centering
\includegraphics[width=0.49\columnwidth]{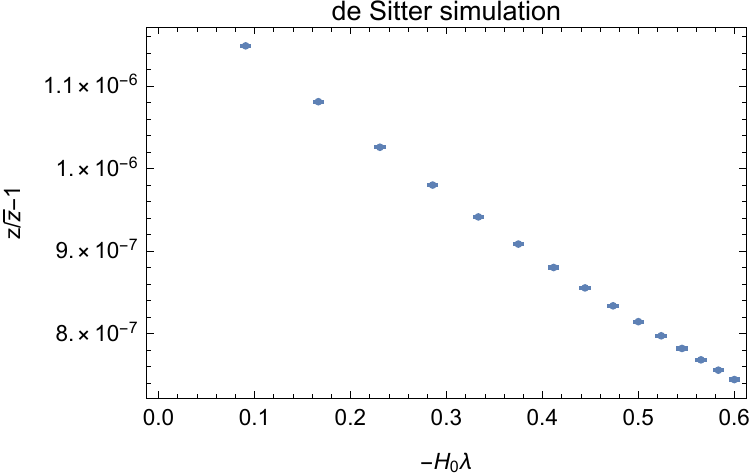}
\hfill
\includegraphics[width=0.49\columnwidth]{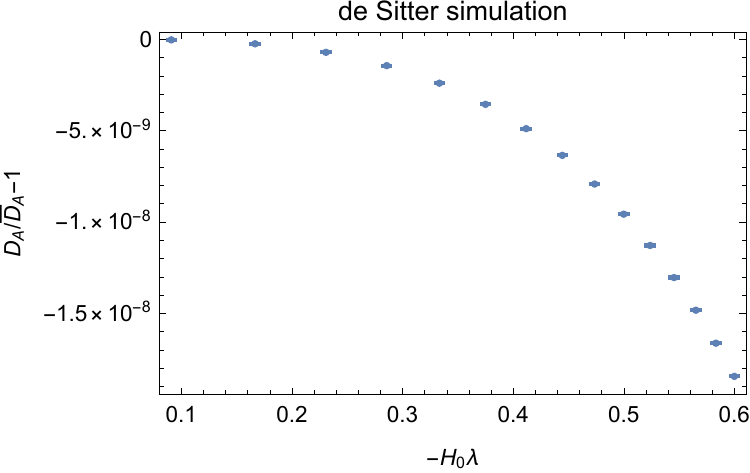}\\
\includegraphics[width=0.49\columnwidth]{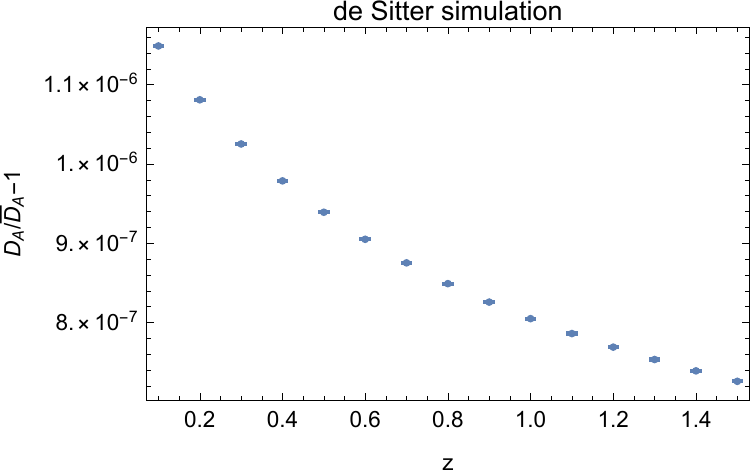}
\hfill
\includegraphics[width=0.49\columnwidth]{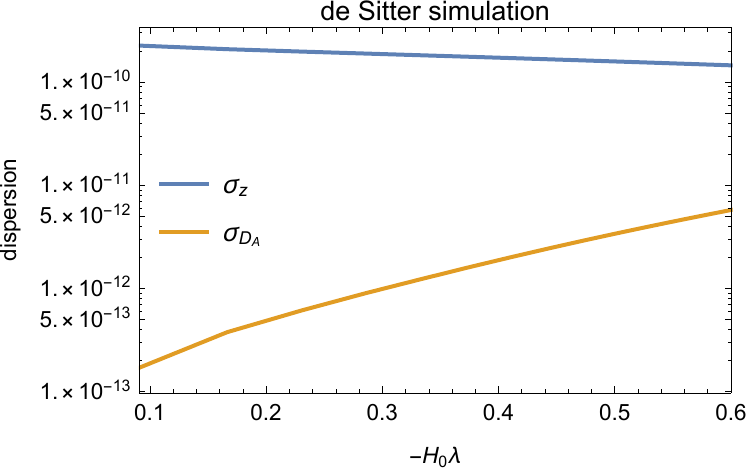}
\caption{Comparison between the output of our ray-tracing code and the known exact results for a de Sitter space-time. \emph{Top-left panel:} The error on the $z(\lambda)$ relation; disks represent averages over $10^5$ light beams shot in random directions, while errors bars (barely visible) indicate the standard deviation of $z(\lambda)$ over the same data set. \emph{Top-right panel:} The same plot for $D\e{A}(\lambda)$. \emph{Bottom-left panel:} The same plot for $D\e{A}(z)$. \emph{Bottom-right panel:} The size of the error bars on the two plots in the top two panels.}
\label{fig:results_dS}
\end{figure}

In order to test the accuracy of our ray-tracing code, we considered the case of a de Sitter space-time, i.e. a universe with no matter but with a non-zero cosmological constant. In this case the potentials in the post-Newtonian metric~\eqref{eq:metric} are given by
\begin{equation}
\Phi = \frac{\Lambda}{6} \, ,
\qquad
\Psi = -\frac{\Lambda}{12} \, .
\end{equation}
We can now simulate observations within this geometry and compare them with the known exact expressions in de Sitter:
\begin{align}
\bar{z}(\lambda) = \frac{-H_0 \lambda}{1+H_0 \lambda} \qquad {\rm and} \qquad
\bar{D}\e{A}(\lambda) = -\lambda \, ,
\end{align}
which combine to give
\begin{align}
\bar{D}\e{A}(z) = \frac{1}{H_0} \frac{z}{1+z} \, ,
\end{align}
where $H_0=\sqrt{\Lambda/3}$. Recall that we chose the wave four-vector~$\vect{k}$ to be future oriented, hence $\lambda\leq 0$ in the past. The relative difference between the output of our ray-tracing code and the exact results above is displayed in figure~\ref{fig:results_dS}. It can be seen from this figure that the accuracy on the $z(\lambda)$ relation in our numerical implementation is at the level of about one part in $10^{6}$, which is even better than our estimates in section~\ref{subsec:code}. The accuracy on the $D\e{A}(\lambda)$ relation is two orders of magnitude better, and also converges as $\sim \Delta t /L$. Finally, the dispersion of the data is much smaller than the mean error. In other words, numerical errors appear like a systematic bias of the data more than a random process. We interpret this fact as being due to the Euler integration, for which the local errors are quadratic, and hence are always cumulative (as they have the same sign). 

\bibliographystyle{JHEP}

\providecommand{\href}[2]{#2}\begingroup\raggedright\endgroup


\end{document}